\documentclass[12pt]{article}
\usepackage{scicite}
\usepackage{url}

\usepackage{times}
\usepackage{graphicx}
\usepackage{physics}

\usepackage{amsthm}
\usepackage{amsmath,bm}
\usepackage{amssymb}

\usepackage{newtxtext,newtxmath}
\usepackage[letterpaper,margin=1in]{geometry}
\usepackage{afterpage}
\usepackage{bbold}

\renewenvironment{abstract}
	{\quotation}
	{\endquotation}
	
\date{}

\makeatletter
\renewcommand{\fnum@figure}{\textbf{Figure \thefigure}}
\renewcommand{\fnum@table}{\textbf{Table \thetable}}
\makeatother

\begin{document}

\title{\bfseries \boldmath Multiparameter estimation with an array of entangled atomic sensors}
\author{
Yifan~Li$^{1}$, 
Lex~Joosten$^{1}$, 
Youcef~Baamara$^{2}$, 
Paolo~Colciaghi$^{1}$, 
Alice~Sinatra$^{2\ast}$, \and
Philipp~Treutlein$^{1\ast}$,
Tilman~Zibold$^{1}$\and
\small$^{1}$Department of Physics, Klingelbergstrasse 82, 4056 Basel, Switzerland.\and
\small$^{2}$Laboratoire Kastler Brossel, ENS-Université PSL, CNRS, Université Sorbonne et Collège de France, \and
\small24 rue Lhomond, 75231 Paris, France.\and
\small$^\ast$Corresponding author. Email: alice.sinatra@lkb.ens.fr, philipp.treutlein@unibas.ch
}

\newcommand{\mean}[1]{\left\langle #1 \right\rangle}
\newcommand{\bb}[1]{\left( #1 \right)}
\newcommand{\Sx}{\hat{S}^x}
\newcommand{\Sy}{\hat{S}^y}
\newcommand{\Sz}{\hat{S}^z}
\newcommand{\hc}{\hat{c}}
\newcommand{\hcd}{\hat{c}^{\dagger}}
\newcommand{\nrel}{n_{\rm rel}}
\newcommand{\V}{\mathcal{V}}
\newcommand{\com}[1]{#1 ^+}
\newcommand{\diff}[1]{#1 ^-}

\newcommand{\Var}{{\rm Var}}
\newcommand{\Cov}{{\rm Cov}}
\newcommand{\Au}{{1\uparrow}}
\newcommand{\Ad}{{1\downarrow}}
\newcommand{\Bu}{{2\uparrow}}
\newcommand{\Bd}{{2\downarrow}}
\newcommand{\Cu}{{3\uparrow}}
\newcommand{\Cd}{{3\downarrow}}

\newcommand{\kAu}{{\ket{1\uparrow}}}
\newcommand{\kAd}{{\ket{1\downarrow}}}
\newcommand{\kBu}{{\ket{2\uparrow}}}
\newcommand{\kBd}{{\ket{2\downarrow}}}
\newcommand{\kCu}{{\ket{3\uparrow}}}
\newcommand{\kCd}{{\ket{3\downarrow}}}

\newcommand{\ncom}{n ^+}
\newcommand{\ndiff}{n ^-}
\newcommand{\Ccom}{C_+}
\newcommand{\Cdiff}{C_-}
\newcommand{\gABdiff}{g^-_{1\rightarrow B}}
\newcommand{\gBAdiff}{g^-_{2\rightarrow A}}
\newcommand{\gABcom}{g^+_{1\rightarrow B}}
\newcommand{\gBAcom}{g^+_{2\rightarrow A}}
\newcommand{\nABdiff}{n^-_{1\rightarrow B}}
\newcommand{\nBAdiff}{n^-_{2\rightarrow A}}
\newcommand{\nABcom}{n^+_{1\rightarrow B}}
\newcommand{\nBAcom}{n^+_{2\rightarrow A}}
\newcommand{\ttheta}{\Tilde{\theta}}
\newcommand{\tTheta}{\Tilde{\Theta}}
\newcommand{\meanSx}{\mean{\hat{S}_x}}
\newcommand{\meanSxA}{\mean{\hat{S}_x^1}}
\newcommand{\meanSxB}{\mean{\hat{S}_x^2}}
\newcommand{\yb}[1]{{\color{red}[(YB) #1]}}

\renewcommand{\figurename}{\textbf{Fig.}}
\renewcommand{\thefigure}{\textbf{\arabic{figure}}}

\maketitle


\begin{abstract}\bfseries \boldmath
In quantum metrology, entangled states of many-particle systems are investigated to enhance measurement precision of the most precise clocks and field sensors. While single-parameter quantum metrology is well established, many metrological tasks require joint multiparameter estimation, which poses new conceptual challenges that have so far only been explored theoretically. 
We experimentally demonstrate multiparameter quantum metrology with an array of entangled atomic ensembles. By splitting a spin-squeezed ensemble, we create an atomic sensor array featuring inter-sensor entanglement that can be flexibly configured to enhance measurement precision of multiple parameters jointly. Using an optimal estimation protocol, we achieve significant gains over the standard quantum limit in key multiparameter estimation tasks, thus grounding the concept of quantum enhancement of field sensor arrays and imaging devices.
\end{abstract}

\noindent
Atomic precision sensors such as atomic clocks \cite{Ludlow2015}, magnetometers \cite{Budker2007}, and inertial sensors \cite{Cronin2009} play an important role in science and technology. Many state-of-the-art devices are limited by the intrinsic quantum noise associated with measurements on a finite number of sensor particles, giving rise to the standard quantum limit (SQL) \cite{Giovannetti2004}. Quantum metrology aims at reducing this noise by harnessing entanglement between the particles \cite{Pezze2018a}, promising significant improvements for sensor applications in fundamental physics and technology \cite{Sinatra2022}. 
Quantum metrology of a single parameter, such as the frequency of an atomic transition or a single component of a magnetic field, has been demonstrated in proof-of-principle experiments \cite{Esteve2008,Appel2009,Schleier-Smith2010,Riedel2010,Gross2010,Leroux2010,Sewell2012} and recently also in metrology-grade setups \cite{Eckner2023,Huang2023a,Robinson2024}. 

Multiparameter estimation is a new frontier in quantum metrology that is receiving great interest 
\cite{Szczykulska2016,Gessner2018,DemkowiczDobrzanski2020,Liu2020,Sidhu2020,Albarelli2020,Pezze2025} because of its relevance for vector field sensors \cite{Kaubruegger2023,Meng2023}, imaging devices \cite{Rehacek2017,Albarelli2020,Baamara2023}, sensor arrays \cite{Gessner2018,Proctor2018,Gessner2020,Gorecki2022,Corgier2023,Malitesta2023,Fadel2023}, and clock networks \cite{Komar2014}. 
While for single-parameter quantum metrology there is a clear theoretical framework \cite{Pezze2018a}, the joint estimation of multiple parameters with quantum sensors is surprisingly complex from a conceptual point of view. For parameters encoded by non-commuting Hamiltonians, the incompatibility of optimal measurements poses a fundamental challenge \cite{Ragy2016,Carollo2019,DemkowiczDobrzanski2020,Kaubruegger2023,Goldberg2021}. In the case of distributed sensing with parameters encoded by commuting Hamiltonians on spatially separated sensor modes, intriguing and intensely debated questions arise regarding the optimal strategy and the possible enhancements provided by inter-sensor entanglement \cite{Gessner2018,Proctor2018,Gessner2020,Gorecki2022,Corgier2023,Malitesta2023,Fadel2023}.
Further challenges arise from constraints on sensor control and detection, and the presence of (possibly correlated) technical noise \cite{Corgier2023,Hamann2024}. 
Due to the complexity of the problem, statements about quantum gain in multiparameter estimation generally depend on the framework adopted. 
While these questions are intensely investigated theoretically, experiments are only beginning to explore this field \cite{Colangelo2017a,Guo2020,Malia2022,Lipka2023}.  

A paradigmatic system for multiparameter quantum sensing is an array of spatially separated atomic ensembles that can be individually controlled and detected \cite{Gessner2018,Gessner2020,Baamara2023,Fadel2023}, such as in an atomic field imaging sensor \cite{Muessel2014,Schaeffner2024} or in an optical lattice clock \cite{Robinson2024}.
The parameters are local spin rotation angles imprinted on the ensembles and the task is to estimate these parameters or certain nonlocal linear combinations of interest. 
Previous experiments demonstrated quantum gain in estimating a single parameter combination with distributed entanglement \cite{Guo2020,Malia2022}.
The scenario considered here is a true multiparameter estimation problem, where each sensor reveals a local parameter value in each experimental run and the question is how entanglement within each ensemble and between the ensembles can enhance measurement precision in multiple parameters jointly. This may require adapting the input state dynamically within the given set of resources \cite{Baamara2023}.
While these questions have recently attracted considerable theoretical interest and different sensing protocols have been proposed \cite{Gessner2018,Gessner2020,Baamara2023,Malitesta2023,Fadel2023}, experimental demonstrations of multiparameter estimation with entangled atomic ensembles are lacking.


In this work, we use an array of atomic Bose-Einstein condensates (BECs) whose collective spins are entangled with each other and can be individually manipulated and detected to demonstrate quantum gain in joint multiparameter estimation of a set of parameters imprinted on the array and their nonlocal linear combinations. Our experiment shows that inter-sensor entanglement enhances the performance of sensor arrays \cite{Gessner2018,Gessner2020} and constitutes an important proof of concept for quantum enhanced field sensors and imaging devices \cite{Baamara2023}.

\subsection*{Joint multiparameter estimation}
\begin{figure}
\centerline{\includegraphics [width = 0.9\textwidth]{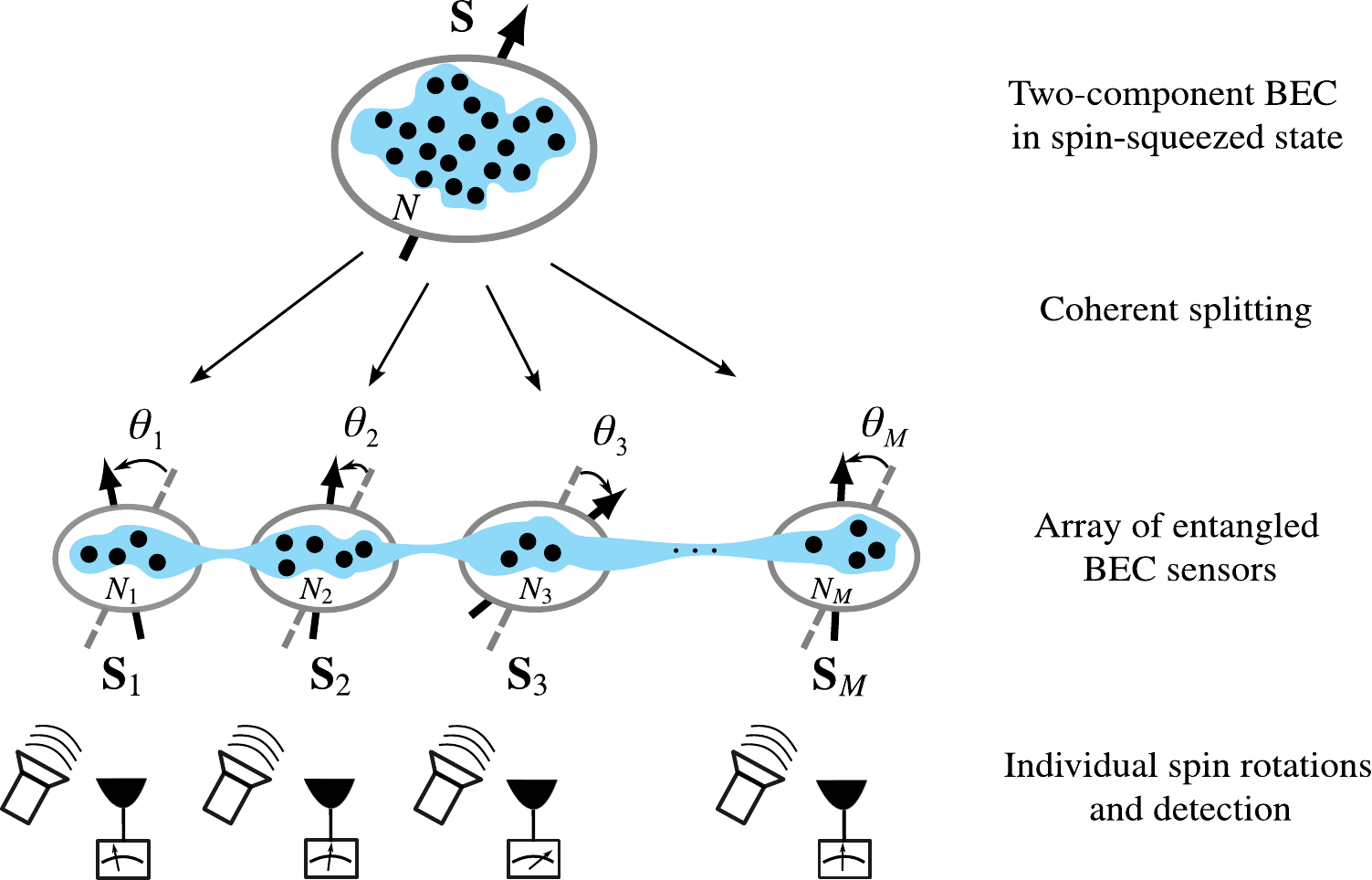}} 
\caption{\label{fig:sensornetwork} \textbf{Array of entangled atomic sensors for multiparameter estimation.} An array of $M$ sensors, each consisting of a collective spin $\mathbf{S}_k$ of $N_k$ two-level atoms, is used to determine $M$ parameters $\theta_1$, $\theta_2$, \ldots , $\theta_M$ that are encoded on the sensors as local spin rotations. The sensor spins are prepared by coherently splitting a two-component BEC in a spin-squeezed state, resulting in entanglement between atoms within each sensor and between different sensors. In combination with individual spin rotations and detection, the entanglement enables a statistical gain in the determination of the $M$ parameters compared to the case without quantum correlations. 
}
\end{figure}

Consider an array of $M$ quantum sensors operating in parallel \cite{Gessner2018,Gessner2020,Baamara2023}, each consisting of an ensemble of $N_k$ two-level atoms that form a collective spin \cite{Pezze2018a} $\mathbf{S}_k$, with $k = 1, \ldots, M$, see Fig.~\ref{fig:sensornetwork}.
The sensors measure a family of unknown parameters $\boldsymbol{\theta}= (\theta_1,\ldots,\theta_M)$ representing e.g.\ the spatial distribution of a field $B(\mathbf{r}_k)$, which is imprinted as local spin rotations by angles $\theta_k \propto B(\mathbf{r}_k)$ on the ensembles. Here, $B$ is a magnetic field, but other quantities such as electric fields or gravity can be imprinted in a similar way through evolution with suitable Hamiltonians.
The sensors are read out by measuring suitable components of each $\mathbf{S}_k$ and the whole experiment is repeated $\mu$ times. 
The goal of multiparameter quantum metrology is to jointly estimate with the highest possible precision the local parameters $\theta_k$ or certain nonlocal combinations $\mathbf{n}\cdot\boldsymbol{\theta} = n_1\theta_1 + \ldots + n_M \theta_M$, where $\mathbf{n}=(n_1,\ldots,n_M)$ is a unit vector of coefficients determining the specific linear combination of interest.
For example, in the case of $M=2$ ensembles, $\mathbf{n}_+ = (1,1)/\sqrt{2}$ gives a measurement of the sum $\theta_+ = (\theta_1 + \theta_2)/\sqrt{2}$ and $\mathbf{n}_- = (1,-1)/\sqrt{2}$ a measurement of the difference $\theta_- = (\theta_1 - \theta_2)/\sqrt{2}$ of the parameters, corresponding to the average field and the field gradient, respectively.

In such a multiparameter estimation task, the optimal management of resources is a complex problem \cite{Gessner2018,Proctor2018,Albarelli2020,Gessner2020,Liu2020,Sidhu2020,Gorecki2022} and the expected performance depends on the scenario considered. In accordance with the experimental constraints and capabilities, we consider the total number of atoms $N=\sum_{k=1}^M N_k$ and the total number of preparations $\mu$ of the system as fixed resources and assume that every sensor can be manipulated and measured individually.
The performance of such a sensor array has been theoretically analyzed \cite{Gessner2018,Gessner2020}, showing that entanglement between the atoms in each ensemble as well as entanglement between the different ensembles can enhance the measurement precision compared to the case of non-entangled atoms. Moreover, it has been shown that entanglement both within and between the ensembles is necessary to achieve the highest performance in estimating a single nonlocal parameter combination \cite{Gessner2018}.

Multiparameter squeezing \cite{Gessner2020}, also called multimode squeezing in other contexts \cite{Guo2020}, is a particularly promising strategy for quantum enhancement in sensor arrays. Similar to spin-squeezing of a single atomic ensemble \cite{kitagawa_1993,Wineland1992}, which has been the most successful approach to quantum metrology with atomic sensors \cite{Pezze2018a}, it is comparatively simple to generate, compatible with standard interferometric sequences and detection methods, and robust against decoherence.
A multiparameter squeezed state features quantum correlations of the sensor spins $\mathbf{S}_k$ that squeeze the noise in the measurement of specific combinations of the parameters $\theta_k$. For nonlocal parameter combinations, this requires nonlocal squeezing in a corresponding superposition of sensor modes. For example, if the whole sensor array is prepared in a squeezed state of the global spin $\mathbf{S}=\sum_{k=1}^M\mathbf{S}_k$ and the atoms are equally distributed, $N_k=N/M$, the linear combination corresponding to the sum $(\theta_1 + \ldots + \theta_M)/\sqrt{M}$ of all parameters can be measured with quantum gain, while all other orthogonal combinations will be measured with a statistical uncertainty greater than that for independent atoms (see Supplementary Text). 
However, as we show theoretically and experimentally, local rotations of the individual sensor spins $\mathbf{S}_k$ can reconfigure the quantum correlations between the sensors, allowing us to achieve quantum enhancement for multiple parameters jointly, using global squeezing of the initial state as the resource \cite{Baamara2023}.
Moreover, a suitable distribution $N_k$ of atoms into the $M$ sensors in combination with local rotations allows us to enhance the measurement of any parameter combination $\mathbf{n}\cdot\boldsymbol{\theta}$ of interest \cite{Gessner2020,Baamara2023} (see Supplementary Text).

\begin{figure}
\centering
\begin{minipage}{1\textwidth}
  \centering
  \includegraphics[width=1\linewidth]{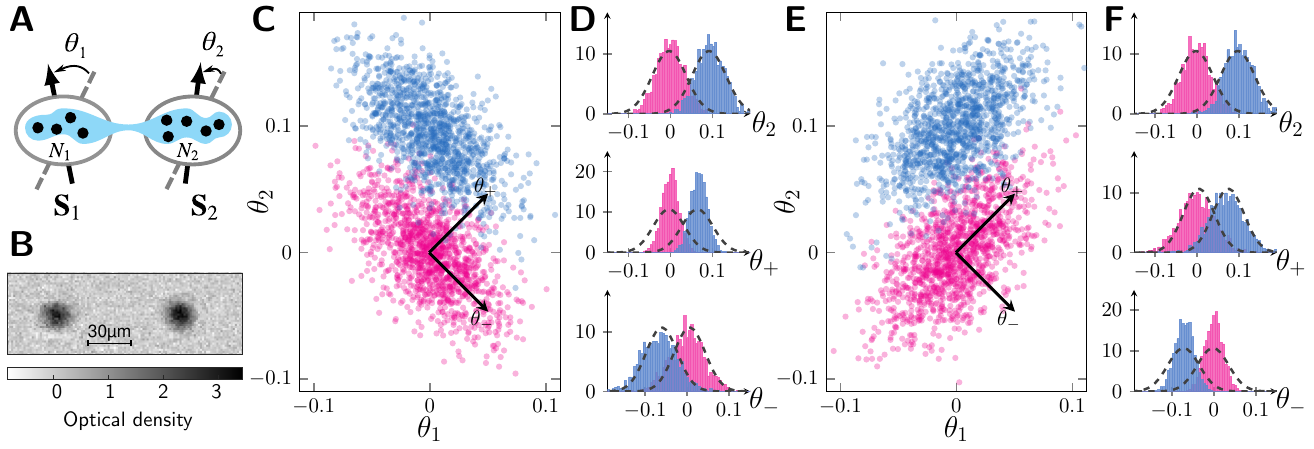}
  \caption{\textbf{Joint estimation of two parameters with two entangled atomic sensors.} (A) Parameters $\theta_1$ and $\theta_2$ are imprinted on the two sensor spins. (B) Absorption image of the two atomic clouds with $N_1\approx N_2$. (C) Correlation plot of simultaneous measurements of $\theta_1$ and $\theta_2$, showing strong correlations due to the inter-sensor entanglement. Two datasets are shown for two different values of $\theta_2$, each with $1200$ repetitions (purple and blue color, respectively). (D) Histograms obtained from the measurements in (C) for $\theta_2$ (top), $\theta_+$ (middle), and $\theta_-$ (bottom). The measurement of $\theta_+$ exploits the inter-sensor entanglement, resulting in the smallest variances. Dashed lines: distribution for an ideal coherent spin state. (E) Correlation plot similar to (C), but for measurements with a $\pi$-pulse applied to $\mathbf{S}_2$ prior to parameter imprinting. (F) Histograms for the data in (E). Now, the measurement of $\theta_-$ shows minimal variance due to the entanglement. }
  \label{fig:correlations}
\end{minipage}%
\end{figure}

\subsection*{Preparation of entangled sensor array} 


In our experiment, the atomic sensor array is realized by spatially splitting a spin-squeezed BEC of $N\approx 1450$ $^{87}$Rb atoms \cite{Riedel2010} into the $M$ ensembles, using coherent splitting techniques similar to ref.~\cite{Colciaghi2023}, which we extend here to enable splitting into more than two ensembles and to adjust the splitting ratios as desired, while maintaining full coherent control (see Supplementary Text).

In each ensemble $k$, the atoms are prepared in a superposition of hyperfine ground states $\ket{k\uparrow}$ and $\ket{k\downarrow}$ that define the collective spin \cite{Pezze2018a} $\mathbf{S}_k$. 
Arbitrary spin rotations can be applied to each sensor individually by coupling the states with resonant microwave and radio frequency magnetic fields.
By absorption detection of the atom numbers $N_{k\uparrow}$ and $N_{k\downarrow}$ in the two states we can directly measure $N_k = N_{k\uparrow} + N_{k\downarrow}$ and the collective spin component $S^z_k=(N_{k\uparrow}-N_{k\downarrow})/2$.  

As a source of entanglement we prepare the initial BEC in a spin-squeezed state of the global spin $\mathbf{S}$, where all atomic spins are entangled with each other \cite{Pezze2018a}. 
Using controlled atomic collisions on an atom chip \cite{Riedel2010}, we prepare states with a Wineland spin-squeezing parameter $\xi^2 = N \Var(S_z)/|\langle S_x \rangle |^2 \approx -6.5(2)$~dB and spin length $\langle S_x\rangle = CN/2$ with contrast $C=0.94(1)$. 
Upon spatial splitting into the sensor clouds (see Supplementary Text), the spin-squeezing results in Einstein-Podolsky-Rosen entanglement between the sensor spins $\mathbf{S}_k$, as illustrated in Fig.~\ref{fig:sensornetwork} and as we have demonstrated previously for two clouds \cite{Colciaghi2023}. Here we extend this technique to multiple ensembles and use it as a resource for multiparameter quantum metrology.

\subsection*{Joint estimation with two entangled sensors}

We first demonstrate joint multiparameter estimation with two entangled atomic sensors, see Fig.~\ref{fig:correlations}. The ensemble is symmetrically split, $N_1\approx N_2 \approx N/2$, and the sensor spins are initially polarized along $S_k^x$. The parameters $\theta_1$ and $\theta_2$ are encoded as small angle rotations of the two sensor spins around the $y$-axis. By measuring the atom numbers in all four states involved, the parameters can be directly estimated as $\theta_k \approx S_k^z / \langle S_k^x \rangle=(N_{k\uparrow}-N_{k\downarrow})/C_k \langle N_k \rangle$. Figure~\ref{fig:correlations}C shows such simultaneous measurements of $\theta_1$ and $\theta_2$, one dataset with and one without a shift applied to $\theta_2$. 
Local entanglement in each ensemble reduces the variance of both $\theta_1$ and $\theta_2$ by $-1.3(2)$~dB below the SQL. However, such an estimation strategy does not exploit the entanglement between the sensors, which manifests itself in strong correlations between the measurement outcomes of $\theta_1$ and $\theta_2$.

To exploit the inter-sensor entanglement, we should estimate the non-local parameter $\theta_+ = (\theta_1 + \theta_2)/\sqrt{2}$, 
which is sensitive to the squeezed global spin component
$S^z=S_1^z+S_2^z$ and can be estimated with a variance of $\Var(\theta_+)=2\xi^2/\mu N$ using $N$ atoms and $\mu$ repetitions of the experiment (see Supplementary Text). Compared to the SQL obtained with unentangled atoms in an ideal coherent spin state, $\Var_\mathrm{SQL}(\theta_\pm)=2/\mu N$, 
we recover 
the full enhancement $\xi^2$ provided by the spin-squeezed state. For the data in Fig.~\ref{fig:correlations}C, we find that $\Var(\theta_+)$ is reduced by $-5.6(2)$~dB below the SQL, which is also evident in the narrow histogram in Fig.~\ref{fig:correlations}D.
The orthogonal linear combination $\theta_- = (\theta_1 - \theta_2)/\sqrt{2}$, on the other hand, which we can also access due to the individual readout of the sensors, is estimated from the same data with $\Var(\theta_-)\approx 2/\mu N C^2$, slightly above the SQL.

Alternatively, we can apply a local $\pi$ rotation to invert the sign of $\mathbf{S}_2$ prior to imprinting the parameters. This transfers the quantum correlations between the sensors into the antisymmetric mode so that $S_1^z-S_2^z$ is squeezed. Now, $\theta_-$ can be estimated with $\Var(\theta_-)=2\xi^2/\mu N$, while $\Var(\theta_+)\approx 2/\mu N C^2$ remains above the SQL. Figures~\ref{fig:correlations}E and \ref{fig:correlations}F show data taken in this way, showing an improvement of $-5.6(2)$~dB below the SQL in $\Var(\theta_-)$.

\begin{figure}
\centering
{\includegraphics[width=\linewidth]{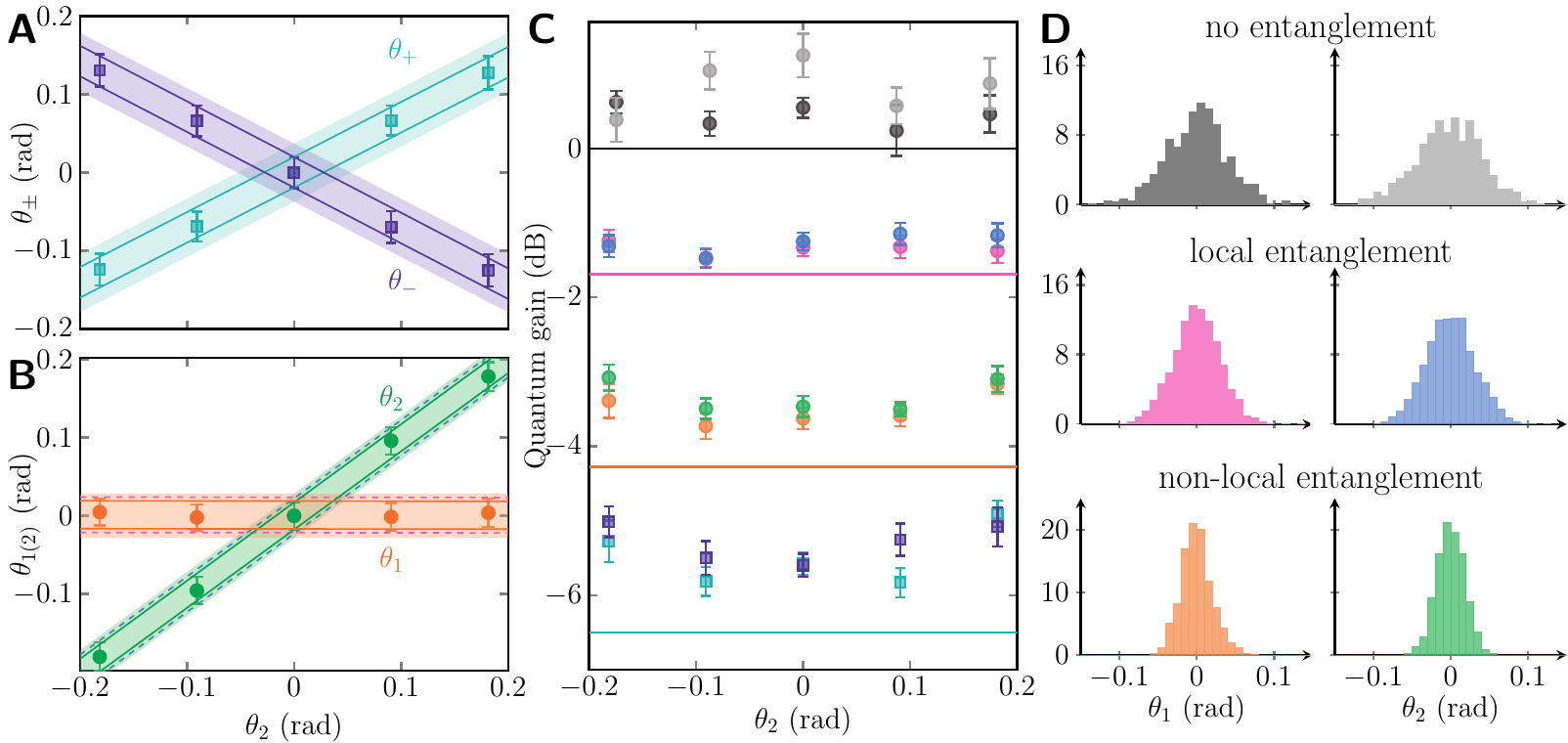}}
\caption{\label{fig:jointestimate} \textbf{Joint estimation of two local parameters enhanced by nonlocal squeezing.} (A) Measurement results of $\theta_+$ (teal) and $\theta_-$ (violet) for different applied rotations $\theta_2$. Error bars: standard deviations (SD) of measurement outcomes. Solid lines: linear fit to SD. Shaded areas: SQL for an ideal coherent spin state. (B) Joint estimation of $\theta_1$ (orange) and $\theta_2$ (green) from properly weighted measurements of $\theta_+$ and $\theta_-$ as described in the text, with error bars and solid lines indicating SD as in (A). Dashed lines: SD obtained if inter-sensor entanglement is ignored. Shaded areas: SQL. (C) Comparison of quantum gains for estimating $\theta_1$ and $\theta_2$ using different strategies: Unentangled atoms (gray and light gray), local measurements ignoring inter-sensor entanglement (pink and blue), and joint estimation using non-local entanglement (orange and green). Solid lines indicate the corresponding theoretical expectations. The square points in teal and violet show the quantum gain for estimating only $\theta_+$ or only $\theta_-$, respectively. Teal line: initial squeezing. All error bars are standard errors of the mean. (D) Histograms of the measurements of $\theta_1$ and $\theta_2$ at an applied $\theta_2=0$ using colors as in C. The top histograms show results for unentangled atoms. The histograms in the middle and bottom rows show data from the same experimental runs with entangled atoms. In the middle row, local estimators make use only of local entanglement within each sensor. In the bottom row, joint estimation makes use also of the non-local entanglement between the sensors.} 
\end{figure}

Our strategy to estimate \textit{both} $\theta_+$ and $\theta_-$ with quantum enhancement, is to alternate between these two settings, performing $\mu/2$ measurements with and $\mu/2$ without the $\pi$ rotation of $\mathbf{S}_2$, respectively, so that the overall resources are unchanged. 
To fully exploit the information provided by both sets of measurements, we estimate both $\theta_+$ and $\theta_-$ in each of the two settings, resulting in four estimates that we combine with appropriate statistical weights (see Supplementary Text). 
This allows us to jointly estimate $\theta_+$ and $\theta_-$, theoretically with identical uncertainties $\Var(\theta_\pm) = \frac{4}{\mu N}\frac{\xi^2}{1+C^2\xi^2}$. With respect to the SQL, the gain here is $\frac{2\xi^2}{1+C^2\xi^2}\approx 2\xi^2$ for $\xi^2 \ll 1$. 
Since the estimators for $(\theta_1,\theta_2)$ are orthonormal linear combinations of $(\theta_+,\theta_-)$, they can be obtained with the same variances  $\Var(\theta_{1,2})=\Var(\theta_\pm)$ from the same dataset. Figure~\ref{fig:jointestimate} shows experimental data for such joint estimation of the local parameters $\theta_1$ and $\theta_2$, using the nonlocal squeezed state as a resource for quantum enhancement. The observed improvement beyond the SQL is $-3.6(2)$~dB for $\theta_1$ and $-3.5(1)$~dB for $\theta_2$, see Fig.~\ref{fig:jointestimate}C. Theoretically, we expect $-4.3(2)$~dB given the initial squeezing of $\xi^2=-6.5(2)$~dB and a contrast of $C=0.94(1)$, in good agreement with the experiment, given that we do not subtract any technical noise (see Supplementary Text). 

The variances $\Var(\theta_{1,2})$ that we obtain with our protocol equal the harmonic average $\bar{\sigma}^H$ of the eigenvalues of the covariance matrix $\textrm{Cov}(\theta_k,\theta_l)$ of the local estimators $\theta_k = S_k^z / \langle S_k^x \rangle$ in the initial, symmetrically split, spin-squeezed state. The harmonic average is smaller than or equal to the arithmetic average corresponding to the trace of the covariance matrix, and we identify it here as the relevant figure of merit (see Supplementary Text). 
For an ideal spin-squeezed state with a large number of atoms, it can be further shown that $\bar{\sigma}^H$ reaches the harmonic average of the eigenvalues of $(\mu {\cal F})^{-1}$, where $\mu$ is the number of system preparations and ${\cal F}$ the multiparameter quantum Fisher information matrix. This shows that this strategy, which is demonstrated here experimentally, is the optimal strategy for the resources at hand (a spin-squeezed state split symmetrically between the sensors), in that it saturates the corresponding Cramer-Rao bound (see Supplementary Text).

\subsection*{Sensing arbitrary parameter combinations}

\begin{figure}[ht]
\centering
{\includegraphics[width=0.8\linewidth]{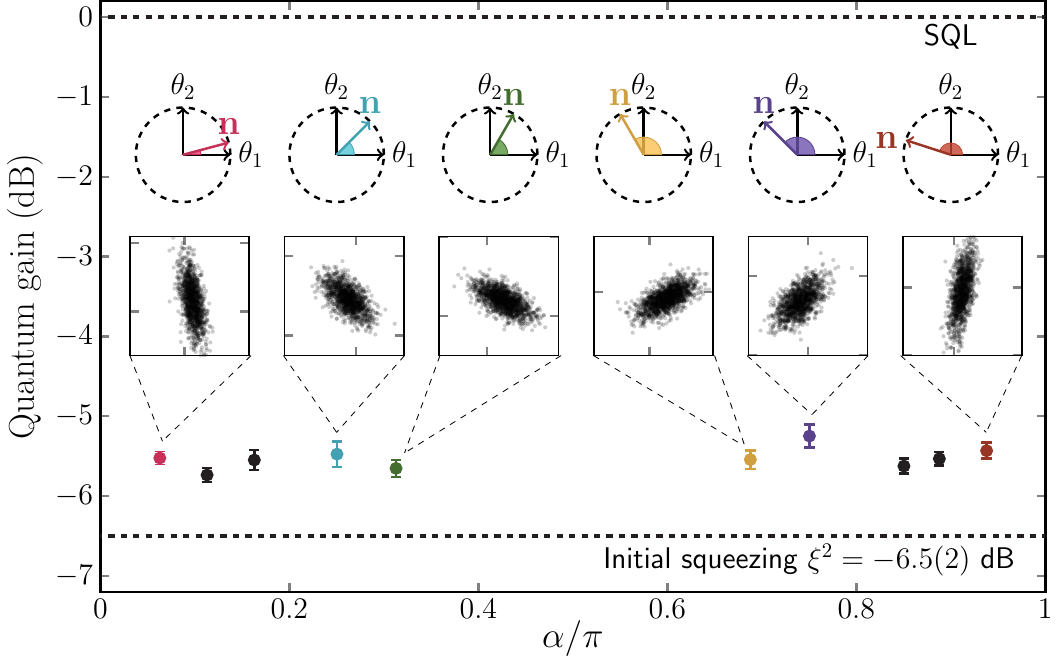}}
\caption{\label{fig:UnequalSplit} 
\textbf{Optimizing the estimation of different linear combinations of parameters.} Linear combinations of two parameters $\mathbf{n}\cdot\boldsymbol{\theta}=\cos (\alpha)\,\theta_1 + \sin (\alpha)\,\theta_2$ are characterized by the mixing angle $\alpha$. For a given $\alpha$, we engineer the quantum correlations between the two sensors in order to harness the entanglement in an optimal way. The resulting quantum gain is shown for ten different linear combinations (data points in main plot). Six examples are shown in the insets, colour code matching the data points.}
\end{figure}

In certain measurement tasks, only a single linear combination of local parameters $\mathbf{n}\cdot\boldsymbol{\theta}$ is of interest, such as in field gradiometry or more generally in measuring a particular multipole moment or Fourier component of a field with a sensor array. In this case, the optimal measurement configuration requires a specific distribution of resources to the sensors \cite{Pezze2024}, which in our case amounts to a particular distribution of local atom numbers $N_k$, see Supplementary Text.  By adjusting the duration of the pulses that are used to split the initial BEC into the sensor clouds, we can adjust this distribution as desired.
We experimentally demonstrate this for the case of two sensors and five different splitting ratios. Together with the ability to invert the signs of the local spins $\mathbf{S}_k$ as described above, this allows us to optimize the measurement configuration for ten different linear combinations $\mathbf{n}\cdot\boldsymbol{\theta}$, see data points in Fig.~\ref{fig:UnequalSplit}. The quantum gain shown in Fig.~\ref{fig:UnequalSplit} quantifies how much better the particular linear combination can be measured compared to a measurement with identical resources but using unentangled atoms in a coherent spin state. 
In the ideal case, this strategy allows exploiting the full enhancement $\xi^2$ provided by the initial spin-squeezed state. In the experiment, we obtain about $-5.5$~dB enhancement for all investigated nonlocal parameter combinations.

\subsection*{More than two entangled sensors}

\begin{figure}[ht]
    \centering \includegraphics[width=\linewidth]{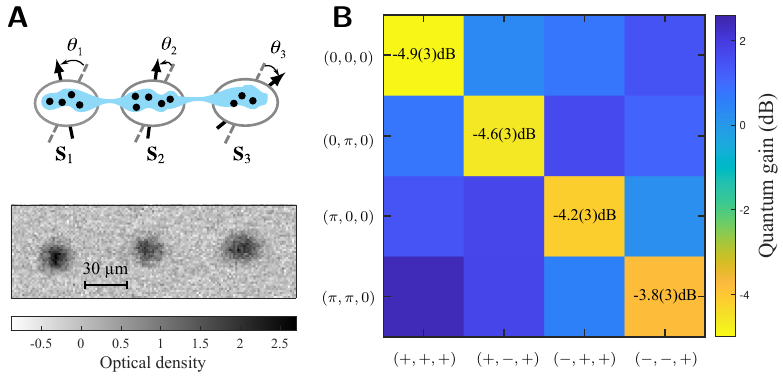}
    \caption{\textbf{Joint multiparameter estimation with $M=3$ entangled atomic sensors.} (A) Schematic of the three entangled sensor spins on which three local parameters are imprinted (top) and absorption image of the three atomic clouds (bottom).
    (B) Matrix of metrological gains compared to the SQL for four different sensor preparations and four estimated parameter combinations. Each row corresponds to a different preparation with $\pi$-pulses applied to the spins $(\mathbf{S}_1,\mathbf{S}_2,\mathbf{S}_3)$ as indicated. Each colum corresponds to the estimation of a different linear combination $(\pm 0.644\, \theta_1\pm 0.431\,\theta_2 + 0.632\,\theta_3)$ with signs $(\pm,\pm,+)$ as indicated. Quantum gain is observed on the diagonal, where the sensor configuration matches the parameter combination.}
    \label{fig:Threeclouds}
\end{figure}

We extend our multiparameter metrology scheme to larger sensor arrays, which raises new conceptual questions on the optimal use of the entanglement in the nonlocal squeezed state. 
For $M$ sensors containing $N_k=N/M$ atoms each, estimating the parameters $(\theta_1$, $\theta_2$, \ldots , $\theta_M)$ with a globally squeezed state, the question arises as to which sensor configurations should be prepared, i.e.\ which of the sensor spins should be subject to $\pi$-rotations prior to parameter imprinting.
We can show that the estimation strategy based on the Hadamard matrix of order $M$, whose elements $\pm 1$ define which sensor spins should be rotated, is optimal (see Ref. \cite{Baamara2023} and Supplementary Text). 
However, Hadamard matrices can only exist for dimensions one, two, and multiples of four.
For other dimensions, we have to resort to a truncated version of the next higher Hadamard matrix, whose rows define the sensor configurations. The simplest case demonstrating this concept is $M=3$, where \textit{four} different configurations of the sensor array have to be prepared to jointly estimate the three local parameters $(\theta_1,\theta_2,\theta_3)$ in an optimal way, corresponding to the rows of a Hadamard matrix of order four with one column truncated. The theoretically expected uncertainty is $\Var(\theta_k)= \frac{M}{\mu N}\cdot\frac{M\xi^2}{1+(M-1)C^2\xi^2}$, where the first factor is the SQL and the second factor the quantum gain (see Supplementary Text).

In Fig.~\ref{fig:Threeclouds} we present data demonstrating joint multiparameter estimation with $M=3$ entangled atomic sensors. 
We split the spin-squeezed BEC into three clouds with $N_1=630(30)$, $N_2=420(20)$, and $N_3 =620(30)$, retaining $-4.9(3)$~dB of squeezing in the global spin $S^z = S^z_1+S^z_2+S^z_3$ after splitting (see Supplementary Text). 
We prepare four different sensor configurations by applying local $\pi$-rotations to the $\mathbf{S}_k$ as indicated in Fig.~\ref{fig:Threeclouds}B. 
For each configuration, we observe a quantum gain of around $-4$~dB beyond the SQL for the linear combination of parameters that matches the sensor configuration, see Fig.~\ref{fig:Threeclouds}B.
For this dataset, these combinations are $(\pm0.644\, \theta_1\pm 0.431\,\theta_2+ 0.632\,\theta_3)$ due to the imbalance in $N_k$ (see Supplementary Text).
Combining the data from all four settings, we can jointly estimate all three local parameters $(\theta_1,\theta_2,\theta_3)$ with quantum gains of $(-1.7(2), -0.8(2), -1.8(2))$~dB beyond the SQL for the same overall resources $N$ and $\mu$. 
If we omit any of the four prepared sensor configurations in the analysis, we obtain lower quantum gains in all three parameters for the same overall $N$ and $\mu$, confirming the quantum advantage of four sensor configurations over three in the case $M=3$.

\subsection*{Outlook}
In this work, we have experimentally demonstrated quantum-enhanced multi-parameter sensing with arrays of up to three atomic sensors. The theoretical analysis shows that our estimation protocol can be extended to an arbitrary number of entangled sensors. While the protocol is optimal for our resources, 
as the number $M$ of jointly estimated parameters increases, the quantum gain for each parameter decreases with $M$, reflecting the fact that only a single collective mode of the array is squeezed in each experimental run but used to enhance all $M$ parameters.

An intriguing perspective for multiparameter estimation with larger sensor arrays is compressed sensing \cite{Baamara2023}, also called multiparameter estimation with nuisance parameters \cite{Suzuki2020,Gross2020}, where one is only interested in a subset ${\cal L}_H$ of all possible nonlocal parameter combinations, with ${\cal L}_H \ll M\ll N$. By specifically preparing sensor configurations that enhance the ${\cal L}_H$ linear combinations of interest, significant quantum gains can be achieved, which is particularly relevant for field imaging and pattern recognition applications \cite{Baamara2023}.

Our experiment marks the first demonstration of multiparameter estimation using globally squeezed states. 
This technique could be transferred to state-of-the-art atomic precision sensors such as optical lattice clocks \cite{Robinson2024}, where entanglement between subensembles in different lattice sites could improve the measurement of gravitational redshifts at short length scales or the characterization of spatially dependent systematic effects.
More generally, our results lay the groundwork for future demonstrations of intriguing sensing schemes such as the entanglement of distant atomic clocks \cite{Polzik2016}, opening up possibilities to study gravitational decoherence \cite{Pikovski2015} and long baseline gravitational wave detection using atom interferometry \cite{Hogan2016}.
Furthermore, our experimental system with collective spins of spatially separated atomic ensembles entangled by one-axis twisting evolution is also well-suited for the realization of recent proposals for vector magnetometry\cite{Kaubruegger2023}, which involve simultaneous sensing of orthogonal magnetic field components for which the Hamiltonians do not commute and the optimal measurements are incompatible. 

\clearpage
\bibliography{scibib} 
\bibliographystyle{sciencemag}

\section*{Acknowledgements}
The authors would like to thank Y.~Castin, R.~Demkowicz-Dobrza{\'n}ski, A.~Hamann, L.~Pezz{\`e}, and P.~Sekatski for helpful discussions. 
\paragraph*{Funding:}
The experiments in Basel were supported by the Swiss National Science Foundation. Y.B. thanks QICS for financial support.

\section*{Supplementary materials}
Supplementary Text\\
Figs. S1 \\
References \textit{(52-56)}

\newpage
\renewcommand{\thefigure}{S\arabic{figure}}
\renewcommand{\thetable}{S\arabic{table}}
\renewcommand{\theequation}{S\arabic{equation}}
\renewcommand{\thepage}{S\arabic{page}}
\setcounter{figure}{0}
\setcounter{table}{0}
\setcounter{equation}{0}
\setcounter{page}{1}

\begin{center}
\section*{Supplementary Materials for\\ 
\textit{Multiparameter estimation with an array of entangled atomic sensors}
}

Yifan~Li, 
Lex~Joosten, 
Youcef~Baamara, 
Paolo~Colciaghi, 
Alice~Sinatra$^{\ast}$,\\
Philipp~Treutlein$^{\ast}$,
Tilman~Zibold\\
\small$^\ast$Corresponding author. Email: alice.sinatra@lkb.ens.fr, philipp.treutlein@unibas.ch
\end{center}

\subsubsection*{This PDF file includes:}
Supplementary Text\\
Figures S1

\newpage

\section{Joint multiparameter estimation with distributed entanglement by equal splitting of an initial spin squeezed state}
We consider a set of $M$ quantum sensors, each consisting of an ensemble of $N_k$ two-level atoms that form a collective spin $\textbf{S}_k$ with $k=1,...,M$. The total ensemble of  $N=\sum_{k=1}^M N_k$ atoms is assumed to be in a spin-squeezed state (SSS) where the global mean spin points in the $x$-direction, and its $z$-component $S_z$ exhibits squeezing characterized by a Wineland spin-squeezing parameter $\xi^2$ \cite{Wineland1992}. These sensors are used to measure an ensemble of $M$ unknown parameters $\bm \theta=(\theta_1,...,\theta_M)^T$, each of which is locally imprinted through a small rotation of the corresponding collective spin around its $y$-axis. 
In our estimation problem, we assume $\mu$ independent preparations of the system, each followed by imprinting the parameters and then measuring the local collective spin components $S_k^z$ of all the sensors $k=1,...,M$.

\subsection{Fixed initial spin-squeezed state}
Performing $\mu$ independent system preparations and local measurements on all sensors, one can estimate each parameter $\theta_k$ as\footnote{Throughout the section, we use the notation $\hat{\theta}_k$ to distinguish between the parameters $\theta_k$ which are fixed and their unbiased estimators $\hat{\theta}_k$ which exhibit fluctuations.} 
$\hat\theta_k=\overline{S}_k^z\big{/}\langle S_k^x\rangle$, where $\overline{S}_k^z=\sum_{i=1}^{\mu}(S_k^z)_i/\mu$ is the statistical average of the measurement results $S_k^z$, and $\langle S_k^x\rangle$ is the expectation value of the observable $S_k^x$ in the quantum state. Here, we distinguish between the statistical average obtained from a finite number of measurements and the expectation value calculated from the quantum state, which assumes an infinite number of measurements, and we assume $\theta_k\ll1 \, \forall k$. The uncertainty on the estimators is quantified by the $M\times M$ covariance matrix $\Sigma^{(\mu)}$ whose elements are $\Sigma^{(\mu)}_{kl}={\rm Cov}(\hat\theta_k,\hat\theta_l)$. In the limit of large $\mu$, it is given by \cite{Gessner2020}
\begin{align}\label{eq: CovMatrix}
\Sigma^{(\mu)}=\frac{1}{\mu}\mathcal{M}^{-1}\qquad\textrm{with}\qquad \mathcal{M}=G^T\Gamma^{-1}G,
\end{align}
where $G_{kl}=-i\langle[S_k^z,S_l^y]\rangle$ and $\Gamma_{kl}=\textrm{Cov}(S_k^z,S_l^z)$ are $M\times M$ matrices. For a symmetric state, as the one resulting from one-axis twisting evolution of an initial coherent spin state, partitioned with an equal number of atoms in each sensor{\footnote{For $N_k\gg1$, and in particular for our experimental parameters, we can neglect the partition noise (see Fig.~2 of reference \cite{Fadel2023}).}}, i.e. $N_k=N/M$ for all $k=1,...,M$, one can show that $G=\left( \langle S_{x}\rangle/M\right) \,\mathbb{1}$, where $\mathbb{1}$ is the $M\times M$ identity matrix. Thus, the covariance matrix (\ref{eq: CovMatrix}) becomes
\begin{align}\label{eq: MatofCov}
\Sigma^{(\mu)} &=\frac{M^2}{\mu|\langle S_{x}\rangle|^2}\Gamma.
\end{align} 
Using both the full symmetry of the initial spin squeezed state under the exchange of two spins and the equal splitting of the total atom number into the $M$ sensors, we can calculate the elements of the matrix $\Gamma$ as   
\begin{align}\label{eq: GamDiag}
\Gamma_{kk}&=\textrm{Var}(S_1^z)\qquad\forall k\\\label{eq: GamHorDiag}
\Gamma_{kl}&=\textrm{Cov}(S_1^z,S_2^z)\qquad\forall k,l\neq k.
\end{align}
Introducing the covariance of any two individual spins-$1/2$ $s_i^z$ and $s_j^z$ in the symmetric spin state, $c_{ij}=\textrm{Cov}(s_1^z,s_2^z)$, it follows that
\begin{align}\label{eq: var}
\textrm{Var}(S_z)=\frac{N}{4}+N(N-1)c_{ij}.
\end{align}
This allows $c_{ij}$ to be expressed as
\begin{align}\label{eq: cij}
c_{ij}=\frac{4\textrm{Var}(S_z)-N}{4N(N-1)}.
\end{align}
Furthermore, one can show that $\textrm{Cov}(S_1^z, S_2^z)=\left(\frac{N}{M}\right)^2c_{ij}$. By using the above equation, we get
\begin{align}\label{eq: covS}
\textrm{Cov}(S_1^z, S_2^z)=\frac{4N\textrm{Var}(S_z)-N^2}{4M^2(N-1)}.
\end{align}
The variance (\ref{eq: var}) can also be written as
\begin{align}\label{eq: varS}
\textrm{Var}(S_z)=M\textrm{Var}(S_1^z)+M(M-1)\textrm{Cov}(S_1^z, S_2^z).
\end{align}
By substituting (\ref{eq: covS}) in (\ref{eq: varS}) we obtain
\begin{align}\label{eq: VarSy1}
\textrm{Var}(S_1^z)=\textrm{Var}(S_z)\left[\frac{1}{M}-\frac{N(M-1)}{M^2(N-1)}\right]+\frac{N^2(M-1)}{4M^2(N-1)}.
\end{align}
By using the equations (\ref{eq: GamDiag}), (\ref{eq: GamHorDiag}), (\ref{eq: covS}) and (\ref{eq: VarSy1}), we can express the elements of the covariance matrix (\ref{eq: MatofCov}) as a function of the initial squeezing $\xi^2=N\mbox{Var}(S_z)\big{/}|\langle S_x \rangle|^2$ and the global mean spin $\langle S_x\rangle$,
\begin{align}\label{eq: SigmaDiag}
\Sigma_{kk}^{(\mu)}&=\frac{1}{\mu(N/M)}\left(\frac{\xi^2}{M}+\frac{N^3(M-1)}{4M(N-1)|\langle S_x\rangle|^2}-\frac{\xi^2(M-1)}{M(N-1)}\right)\qquad\forall k,\\\label{eq: SigmaHorDiag}
\Sigma_{kl}^{(\mu)}&=\frac{\xi^2}{\mu(N-1)}-\frac{N^2}{4\mu(N-1)|\langle S_{x}\rangle|^2}\qquad\forall k,l\neq k.
\end{align}
Equation (\ref{eq: SigmaDiag}) can be rewritten as 
\begin{align}
\label{eq: GainEq}
\frac{{\rm Var}(\hat\theta_k)}{{\rm Var}_{\rm SQL}(\hat\theta_k)}=\frac{\xi^2}{M}+\frac{N^3(M-1)}{4M(N-1)|\langle S_x\rangle|^2}-\frac{\xi^2(M-1)}{M(N-1)}
\end{align}
with the standard quantum limit (SQL) 
\begin{align}
 {\rm Var}_{\rm SQL}(\hat\theta_k) = \frac{1}{\mu(N/M)}.
\end{align}
Equation (\ref{eq: GainEq}) shows that residual local entanglement in each ensemble can in principle reduce the variance of all estimators $\hat\theta_k$ below the SQL by a factor $(M-1)/M$ in the limit $N\gg1$ and $\xi^2\to 0$.
While this estimation strategy shows a small quantum gain, it does not make use of the entanglement between the sensors. 

To exploit the entanglement between the sensors, we need to make use of the covariances in Eq.~(\ref{eq: SigmaHorDiag}) as well. 
The symmetric structure of the covariance matrix $\Sigma^{(\mu)}$ leads to a simple spectrum with two distinct eigenvalues: $\sigma_{\rm min} = \Sigma_{kk}^{(\mu)} + (M-1) \Sigma_{kl}^{(\mu)}$, with multiplicity $g = 1$, and $\sigma_{\rm max} = \Sigma_{kk}^{(\mu)} - \Sigma_{kl}^{(\mu)}$, with multiplicity $g = M-1$. This spectrum reveals that entanglement between the sensors allows one to estimate the linear combination $\theta_{\rm sq}=\bm n_{\rm sq}\cdot\bm\theta$, where $\bm n_{\rm sq}=(1,...,1)^T/\sqrt M$ is the unit eigenvector of $\Sigma^{(\mu)}$ corresponding to the eigenvalue $\sigma_{\rm min}$, with full quantum gain
\begin{align}
\frac{{\rm Var}(\hat\theta_{\rm sq})}{{\rm Var}_{\rm SQL}(\hat\theta_{\rm sq})}=\xi^2,
\end{align}
while all other $M-1$ linear combinations of parameters $\theta_{\rm asq}$ that are orthorgonal to $\theta_{\rm sq}$ are estimated with an estimator variance above the SQL,
\begin{align}
\frac{\textrm{Var}(\hat\theta_{\rm asq})}{{\rm Var}_{\rm SQL}(\hat\theta_{\rm asq})}=\frac{N^3}{4(N-1)|\langle S_x\rangle|^2}-\frac{\xi^2}{N-1} \: > \: 1 \quad \mbox{for} \quad \xi^2 < 1 \,.
\end{align}
However, as discussed in the next section, local control over the sensor spins allows one to change the optimal linear combination of parameters $\theta_{\rm sq}$ whose estimation is enhanced by entanglement between the sensors.

\subsection{Redistribution of the squeezing in different combinations of the local collective spins}
\label{sub:redistribution}
It was theoretically shown in ref.~\cite{Baamara2023} that one can change the optimally estimated parameter combination $\theta_{\rm sq}$ by applying local $\pi$ rotations to the sensor spins prior to imprinting the parameters. In particular, the linear combination 
$\theta_{\rm sq}=\bm n_{\rm sq}\cdot{\bm \theta}$, with $\bm n_{\rm sq}=(\epsilon_1,...,\epsilon_M)^T/\sqrt M$, where $\epsilon_k=\pm1$, can be estimated with the full quantum enhancement by performing a local $\pi$ rotation to the $k^{\rm th}$ sensor whenever $\epsilon_k=-1$. A complete set of orthogonal independent linear combinations of the parameters of this form is provided by the Hadamard transformation when $M=2^p$ with $p$ an integer \cite{Kunz1979}, and in this section, we restrict ourselves to this case. By adapting the choices of $\epsilon_k$ to measure all the $M$ Hadamard coefficients of the signal $\bm \theta$, each using $\mu/M$ independent realizations, and then using the inverse Hadamard transformation, one can infer the original parameters $\bm \theta$ with the same estimator variance for each parameter \cite{Baamara2023}
\begin{align}\label{eq: Good}
\textrm{Var}(\hat\theta_k^{\rm sq})=\frac{\xi^2}{(\mu/M)(N/M)}.
\end{align}
One can further improve on this result by fully exploiting the information from local readout, using also the estimates of the Hadamard coefficients with suboptimal variance. This allows us, in addition to the estimation with the variance (\ref{eq: Good}), to estimate each parameter $M-1$ times with the variance
\begin{align}\label{eq: Bad}
\textrm{Var}(\hat\theta_k^{\rm asq})=\frac{1}{(\mu/M)(N/M)}\left(\frac{N^3}{4(N-1)|\langle S_x \rangle|^2}-\frac{\xi^2}{N-1}\right).
\end{align}
Next, we determine the best way to combine these $M$ estimated values $\hat\theta_k^{\rm sq}$ and $(\hat\theta_k^{{\rm asq},1},...,\hat\theta_k^{{\rm asq},M-1})$ of $\theta_k$, where $\hat\theta_k^{\rm sq}$ is the ``squeezed" estimation of $\theta_k$, and $\hat\theta_k^{{\rm asq},l}$, for $l\in[1,M-1]$, are the $M-1$ ``anti-squeezed" estimations of $\theta_k$. For this, we construct the unbiased estimator
\begin{align}\label{eq: Estimator}
\hat{\theta}_k=\left(\frac{1+\alpha}{2}\right)\hat{\theta}_k^{\rm sq}+\frac{1}{M-1}\left(\frac{1-\alpha}{2}\right)\sum_{l=1}^{M-1} \hat{\theta}_k^{{\rm asq},l}.
\end{align}
Due to the orthogonality of the Hadamard coefficients and the symmetry of the state, one can show that $\hat\theta_k^{\rm sq}$ and each $\hat\theta_k^{{\rm asq},l}$ for $l=1,...,M-1$ are independent. The variance of this estimator can thus be written as
\begin{align}\label{eq: VarEst}
\textrm{Var}(\hat{\theta}_k)=\left(\frac{1+\alpha}{2}\right)^2\textrm{Var}(\hat\theta_k^{\rm sq})+\frac{1}{M-1}\left(\frac{1-\alpha}{2}\right)^2\textrm{Var}(\hat\theta_k^{\rm asq}).
\end{align}
By minimizing the variance (\ref{eq: VarEst}) over $\alpha$, we find that choosing
\begin{align}\label{eq: alphamin}
\alpha=\alpha_{\rm min}=\frac{\textrm{Var}(\hat\theta_k^{\rm asq})/(M-1)-\textrm{Var}(\hat\theta_k^{\rm sq})}{\textrm{Var}(\hat\theta_k^{\rm asq})
/(M-1)+\textrm{Var}(\hat\theta_k^{\rm sq})}
\end{align}
in equation (\ref{eq: Estimator}) gives
\begin{align}\label{eq: stat-enh}
\textrm{Var}(\hat{\theta}_k)=\frac{1}{1/\textrm{Var}(\hat\theta_k^{\rm sq})+(M-1)/\textrm{Var}(\hat\theta_k^{\rm asq})}.
\end{align}
Substituting (\ref{eq: Good}) and (\ref{eq: Bad}) in (\ref{eq: stat-enh}) yields
\begin{align}\label{eq: VarAll}
{\rm Var}(\hat{\theta}_k)=\frac{1}{\mu(N/M)}\left(\frac{M}{\frac{1}{\xi^2}+\frac{M-1}{\frac{N^3}{4(N-1)|\langle S_x\rangle|^2}-\frac{\xi^2}{N-1}}}\right) \:\:\: \stackrel{N\gg1}{\simeq}  \:\:\: \frac{1}{\mu(N/M)}\left(\frac{M\xi^2}
{1+(M-1)C^2\xi^2} \right), 
\end{align}
where $C$ is the contrast
\begin{equation}
C=\frac{|\langle S_x\rangle|}{N/2} \,.
\end{equation}
Relative to the SQL, for which ${\rm Var}_{\rm SQL}(\hat{\theta}_k)=1/\left(\mu N/M\right)$, this strategy allows us to jointly estimate all the parameters with the same quantum enhancement
\begin{align}\label{eq: GainAll}
{\rm Var}(\hat{\theta}_k)/{\rm Var}_{\rm SQL}(\hat{\theta}_k) \:\:\: \stackrel{N\gg1}{\simeq}  \:\:\: \frac{M\xi^2}
{1+(M-1)C^2\xi^2} .
\end{align}
In the limit of a large amount of initial squeezing $\xi^2\ll1$ with a fixed number of sensors $M$, equation (\ref{eq: GainAll}) shows that the quantum enhancement, with respect to the SQL, achieved by this strategy is
\begin{align}\label{eq: GainAllLimit}
{\rm Var}(\hat{\theta}_k)/{\rm Var}_{\rm SQL}(\hat{\theta}_k) \:\:\: \stackrel{ M\xi^2 \ll 1}{\simeq}  \:\:\:  M\xi^2.
\end{align}
Note that equation \eqref{eq: VarAll} for ${\rm Var}(\hat{\theta}_k)$, which gives the final variance of the estimators within our protocol, is equal to the harmonic average of the eigenvalues ($\sigma_{\rm min}$ and $\sigma_{\rm max}$ with $(M-1)$ multiplicity) of the covariance matrix $\Sigma^{(\mu)}$ (equation (S1)) of the local estimators $\hat{\theta}_k = S_k^z / \langle S_k^x \rangle$ in the initial spin-squeezed state.

\subsection{Optimality of the joint estimation of $M$ parameters}

\subsubsection{A Cramer-Rao inequality for joint multiparameter estimation}
For multiparameter estimation using a fixed quantum state, a Cramer-Rao inequality was proven \cite{Liu2020},
\begin{equation}
\Sigma^{(\mu)} \ge \left(\mu{\cal F}\right)^{-1},
\label{eq:CR}
\end{equation}
where $\Sigma^{(\mu)}$ is the covariance matrix of the parameter estimators  and ${\cal F}$ is the multiparameter quantum Fisher information matrix in the considered quantum state.
However, as we have pointed out previously, using a single fixed quantum state is not the best strategy to obtain quantum gain in the {\it joint} estimation of all the parameters. 
For this reason, in our protocol, starting from a total number $\mu$ of preparations of a given quantum state (our resource), we change the state of different groups of realizations to reshape their quantum correlations, which allows us to construct estimators of the local parameters displaying jointly a reduced variance with respect to the standard quantum limit.

In this new scenario, we have to adapt the Cramer-Rao inequality (\ref{eq:CR}). Using the fact that the transformations performed on the initial squeezed state do not change the spectrum of $\Sigma^{(\mu)}$ and ${\cal F}$, we obtain a new Cramer-Rao inequality involving the spectrum of the two matrices (instead of the matrices themselves), or more precisely the harmonic averages 
 $\bar{\sigma}^H$ and $\bar{\lambda}^H$ of their eigenvalues,
\begin{eqnarray}
\bar{\sigma}^H &\equiv& \frac{M}{\sum_{i=1}^M \frac{1}{\sigma_i}} \quad \mbox{where $\sigma_i$ are the eigenvalues of $\Sigma^{(\mu)}$}, \\
\bar{\lambda}^H &\equiv& \frac{M}{\sum_{i=1}^M \frac{1}{\lambda_i}} \quad \mbox{where $\lambda_i$ are the eigenvalues of $(\mu {\cal F})^{-1}$}, 
\end{eqnarray}
that takes the form\footnote{For each type of state (set of pulses) we prepare, the Cramer Rao inequality (\ref{eq:CR}) holds.
The two matrices in (\ref{eq:CR}) are diagonalized by the same orthogonal transformation $A \to {}^TP A P$, where
${}^TP=P=H$ is the Hadamard matrix that connects the parameters $\bm{\theta}$ to the orthogonal Hadamard combinations. We then have:
\begin{equation*}
\left( H \Sigma H \right) \ge \left( H {\cal F}^{-1} H \right).
\label{eq:CRD}
\end{equation*}
This last inequality that is now between diagonal matrices, holds for the eigenvalues
$\sigma_i \ge \lambda_i$ for $i=1,\ldots,M$.  
Since the harmonic average $\bar{a}^H$ is a monotonic function of its arguments $a_i$:
\begin{equation*}
\frac{\partial \bar{a}^H}{\partial a_i} = \frac{\partial}{\partial a_i}  \left( \frac{M}{\sum_{i=1}^M\frac{1}{a_i}} \right) = \frac{M}
{\left(\sum_{i=1}^M \frac{1}{a_i}\right)^2} \frac{1}{a_i^2} > 0 \,,
\end{equation*}   
the inequality between eigenvalues transfers to the harmonic average, and we recover equation (\ref{eq:CRH}).}
\begin{eqnarray}
\boxed{\bar{\sigma}^H  \:\: \ge \:\: \bar{\lambda}^H.}
\label{eq:CRH}
\end{eqnarray}

\subsubsection{Approximate saturation of the Cramer-Rao inequality}
In this subsection, we demonstrate that for an ideal spin squeezed state in the limit of large $N$, the joint estimation protocol detailed in the previous section, for which $\mbox{Var}(\hat{\theta}_k)=\bar{\sigma}^H$, saturates the Cramér-Rao inequality (\ref{eq:CRH}).
Specifically, we show that the variance of the parameter estimators $\hat{\theta}_k$ defined by the equation (\ref{eq: Estimator}), which is the same for each parameter and equals the harmonic average of the eigenvalues of $\Sigma^{(\mu)}$, matches the harmonic average of the eigenvalues of $\left( \mu {\cal F} \right)^{-1}$ in the limit of a pure one-axis twisting spin squeezed state with $N \gg 1$: 
\begin{equation}
\mbox{Var}(\hat{\theta}_k)=\bar{\sigma}^H \simeq \bar{\lambda}^H\,.
\label{eq:CRsat}
\end{equation}

Let us first concentrate on the left hand side of the inequality (\ref{eq:CRH}).
For an ideal spin squeezed state with $N\gg1$, up to the best squeezing time, one has $\langle S_x \rangle \simeq N/2$ and $C \simeq 1$ (almost perfect contrast), so that the variance of the estimator can be approximated by
\begin{align}
{\rm Var}(\hat{\theta}_k)=\frac{1}{\mu(N/M)}\left(\frac{M}{\frac{1}{\xi^2}+\frac{M-1}{\frac{N^3}{4(N-1)|\langle S_x\rangle|^2}-\frac{\xi^2}{N-1}}}\right) \:\:\: \stackrel{C\simeq 1 \, ; \,  N\gg1}{\simeq} \:\:\:
\frac{1}{\mu(N/M)}\left(\frac{M}{\frac{N/4}{\textrm{Var}(S_z)}+M-1}\right).
\label{eq:sH}
\end{align}

Concerning the right hand side of equation (\ref{eq:CRH}), for a pure state, the multiparameter quantum Fisher information matrix ${\cal F}$ is proportional to the covariance matrix of the local generators of rotations that are here the $y$ components of the local collective spins,
\begin{equation}
{\cal F}_{kl}=4 \, \mbox{Cov}(S_k^y,S_l^y).
\end{equation}
The elements of ${\cal F}$ can be deduced from equations (\ref{eq: covS}) and (\ref{eq: VarSy1}) of the previous section by changing $S_z \to S_y$, and the eigenvalues of ${\cal F}$ are
\begin{eqnarray}
\lambda_{\rm sq}^{\cal F} &=& \frac{N}{M} \frac{\textrm{Var}(S_y)}{N/4} \quad \quad \mbox{non degenerate}, \label{eq:Fsq}\\
\lambda_{\rm asq}^{\cal F} &=& \frac{N}{M}\frac{1}{N-1}\left[N- \frac{\textrm{Var}(S_y)}{N/4}\right] \quad \mbox{degenerate}
\times (M-1). \label{eq:Fns}
\end{eqnarray}
For times shorter than the best squeezing time $N^{-1} \ll \chi t \ll N^{-2/3}$ where the system already exhibits important squeezing and the Wigner distribution, representing the fluctuations of the collective spin in the transverse plane, is approximately a minimum uncertainty Gaussian \cite{kitagawa_1993,Sinatra2012},
\begin{eqnarray}
 \langle S_x \rangle &\simeq& \frac{N}{2}, \\
 \Delta S_y \Delta S_z &\simeq&  \frac{N}{4} \quad \Rightarrow \quad \Delta S_y \simeq  \frac{N}{4}\frac{1}{\Delta S_z}\,.
\end{eqnarray}
Inserting these equations into (\ref{eq:Fsq}) and (\ref{eq:Fns}), and using the assumption $N \gg 1$, one gets
\begin{eqnarray}
\lambda_{\rm sq}^{\cal F} & \simeq & \frac{N}{M} \frac{N/4}{\mbox{Var}(S_z)},\\
\lambda_{\rm asq}^{\cal F} & \simeq &  \frac{N}{M}. 
\end{eqnarray}
We can then calculate the harmonic average of the eigenvalues of the matrix $(\mu {\cal F})^{-1}$, to obtain
\begin{equation}
\bar{\lambda}^{H}=\frac{1}{\mu}\frac{M}{\lambda_{\rm sq}^{\cal F}+(M-1)\lambda_{\rm asq}^{\cal F}} \simeq
\frac{1}{\mu(N/M)}\left(\frac{M}{\frac{N/4}{\textrm{Var}(S_z)}+M-1}\right)\,.
\label{eq:lH}
\end{equation}
The comparison of equations (\ref{eq:lH}) and (\ref{eq:sH}) gives the result (\ref{eq:CRsat}).
This shows the optimality of the joint estimation strategy, in the case of an ideal spin squeezed state as a resource, equal splitting of the squeezed ensemble over the $M$ sensors, and a number of sensors that is equal to a power of two, that is $M=2^p$ with $p$ integer.

\subsection{Case of $M=3$ sensors}
Let us now consider the scenario where we have $M=3$ sensor spins, each containing $N/3$ atoms. Since the Hadamard transformation does not exist for $M=3$, we have to treat this case separately. The approach that we follow here is to use the Hadamard transformation for the $M=4$ case, where the fourth local parameter is set to zero. In order to estimate the \textit{three} local parameters $\bm\theta=(\theta_1,\theta_2,\theta_3)^T$, we thus prepare and measure \textit{four} linear combinations:
\begin{align}\label{eq: c1}
c_1&=\frac{1}{\sqrt{3}}(\theta_1+\theta_2+\theta_3),\\
c_2&=\frac{1}{\sqrt{3}}(\theta_1-\theta_2+\theta_3),\\
c_3&=\frac{1}{\sqrt{3}}(\theta_1+\theta_2-\theta_3),\\\label{eq: c4}
c_4&=\frac{1}{\sqrt{3}}(\theta_1-\theta_2-\theta_3),
\end{align}
from which we deduce the local parameters. According to subsection \ref{sub:redistribution}, any of the linear combinations (\ref{eq: c1})-(\ref{eq: c4}) can be chosen to be estimated with a quantum gain. With a fixed total number of system preparations $\mu$, the four linear combinations can then be estimated with a variance
\begin{align}
\textrm{Var}(\hat{c}_{j})=\frac{\xi^2}{(\mu/4)(N/3)}\quad\forall j\in\{1,2,3,4\}.
\end{align}
Comparing this to the corresponding variance using $\mu$ system preparations with a coherent spin state (i.e.~$\xi^2=1$), the quantum gain on the estimation of each linear combination $c_j$ is $4\xi^2$. When a particular linear combination $c_j$ is estimated with quantum gain, the other three linear combinations $c_k$ are estimated with a
variance
\begin{align}
\textrm{Var}(\hat{c}_{j})=\frac{1}{(\mu/4)(N/3)}\left(\frac{2N^3}{9(N-1)|\langle S_x\rangle|^2}+\frac{(N-9)\xi^2}{9(N-1)}\right).
\end{align}
The local parameters can then be inferred once using the four squeezed linear combinations with variance
\begin{align}
\textrm{Var}(\hat{\theta}_k^{\rm sq})=\frac{\xi^2}{(\mu/3)(N/3)},
\end{align}
and three times using the four non-squeezed linear combinations, with a variance
\begin{align}
\textrm{Var}(\hat{\theta}_k^{{\rm nsq},l})=\frac{1}{(\mu/3)(N/3)}\left(\frac{2N^3}{9(N-1)|\langle S_x\rangle|^2}+\frac{(N-9)\xi^2}{9(N-1)}\right).
\end{align}
Here, each local parameter is inferred from four linear combinations that are measured independently. For a given local parameter $k$, all measurement results can be used to construct the unbiased estimator
\begin{align}\label{eq: Estimator_3Clouds}
\hat{\theta}_k=\left(\frac{1+\alpha}{2}\right)\hat{\theta}_k^{\rm sq}+\frac{1}{3}\left(\frac{1-\alpha}{2}\right)\sum_{l=1}^3 \hat{\theta}_k^{{\rm nsq},l}.
\end{align}
By taking into account the covariances between the estimators $\hat{\theta}_k^{\rm sq}$ and $\hat{\theta}_k^{{\rm nsq},l}$ with $l=1,2,3$, which originate from the covariances between the linear combinations measured in the same realization, the minimization of the variance of (\ref{eq: Estimator_3Clouds}) over $\alpha$ yields 
\begin{align}
\alpha_{\rm min} = -\frac{\Sigma_{11}^{(\mu/4)}+4\Sigma_{12}^{(\mu/4)}}{2(\Sigma_{11}^{(\mu/4)}+\Sigma_{12}^{(\mu/4)})},
\end{align}
where $\Sigma_{11}^{(\mu)}$ and $\Sigma_{12}^{(\mu)}$ are given by (\ref{eq: SigmaDiag}) and (\ref{eq: SigmaHorDiag}) respectively. By substituting $\alpha_{\rm min}$ into (\ref{eq: Estimator_3Clouds}) and calculating its variance, each local parameter $\theta_k$ can then be estimated with the same variance
\begin{align}\label{eq: VarAll3Clouds}
{\rm Var}(\hat{\theta}_k)=\frac{1}{\mu(N/3)}\left(\frac{3}{\frac{1}{\xi^2}+\frac{2}{\frac{N^3}{4(N-1)|\langle S_x\rangle|^2}-\frac{\xi^2}{N-1}}}\right) \:\:\: \stackrel{N\gg1}{\simeq}  \:\:\: \frac{1}{\mu(N/3)}\left(\frac{3\xi^2}
{1+2C^2\xi^2} \right).
\end{align}
Note that this last equation coincides with (\ref{eq: VarAll}) when $M=3$.

\section{Measurement of a generic combination of parameters by suitable splitting of an initial spin squeezed state}

As in the previous section, we consider as a resource a fixed total number $\mu$ of preparations of an initially spin-squeezed state with $N$ atoms, where the global mean spin points in the $x$ direction, and its $z$ component $S_z$ exhibits squeezing characterized by a Wineland spin-squeezing parameter $\xi^2$. The difference in this section is that now, in addition to the possibility to flip the spins and to perform local measurements, we add flexibility in distributing the spin-squeezed atoms among the $M$ sensors. 

In this frame,
given a linear combination ${\cal C}$ of the $M$ parameters with real coefficients,
\begin{equation}\label{eq: Lin_Comb}
{\cal C} \equiv \sum_{k=1}^M c_k \theta_k \quad ; \quad c_k \in \mathbb{R},
\end{equation}
we aim to determine the optimal strategy to estimate ${\cal C}$ and its corresponding quantum gain. Note that contrary to the previous sections, we do not impose any normalization condition on the combination ${\cal C}$.

\subsection{Distributed entanglement strategy}
The optimal measurement configuration requires specific distributions of atoms into the $M$ sensors. The one we consider in this subsection is to choose the number of atoms $N_k$ in each sensor to be proportional to $|c_k|$, with the correct normalization factor ${\cal A}$,
\begin{equation}\label{eq: Norm_A}
N_k= {\cal A} |c_k| \quad \mbox{with}  \quad {\cal A}=\frac{N}{\sum_j |c_j| } \quad \mbox{so that} \quad \sum_{k=1}^M N_k = N,
\end{equation}
and we use $\pi$ pulses to introduce the signs of the coefficients $\epsilon_k=c_k/|c_k|$ in front of the $\theta_k$. To first order in $\theta_k$, and for an ideal spin squeezed state with $N\gg1$ where $\langle S_x \rangle \simeq N/2$ and $\langle S^x_k \rangle \simeq N_k/2$, a measurement of the $z$ component of the global spin $S_z$ allows the estimation of the desired linear combination ${\cal C}$,
\begin{equation}\label{eq: Sy_C}
\langle S_z \rangle =\sum_{k=1}^M \langle S^z_k \rangle = \sum_{k=1}^M  \langle S^x_k \rangle \epsilon_k \theta_k  = \sum_{k=1}^M \frac{N_k}{2} \epsilon_k  \theta_k =  \frac{\cal A}{2} \sum_{k=1}^M |c_k| \epsilon_k \theta_k=\frac{\cal A}{2} \sum_{k=1}^M c_k \theta_k=\frac{\cal A}{2}  {\cal C} \,.
\end{equation}

We now calculate the uncertainty in the estimated linear combination $\mbox{Var}(\hat{\cal C})_{\rm SSS}$, achieved by measuring the collective spin observable $S_z$. By definition we have
\begin{align}
\mbox{Var}(\hat{\cal C})_{\rm SSS}\equiv \frac{1}{\mu}\left.\frac{\textrm{Var}(S_z)}
{|\partial_{\cal C}\langle S_z\rangle|^2}\right|_{\theta=0}.
\end{align}
Using (\ref{eq: Sy_C}) then (\ref{eq: Norm_A}), one can rewrite this last equation as
\begin{align}
\mbox{Var}(\hat{\cal C})_{\rm SSS}=\left.\frac{1}{\mu}\frac{N}{N}\frac{\textrm{Var}(S_z)}{\left(\frac{ {\cal A}}
{2}\frac{ \langle S_x\rangle}{\langle S_x\rangle}\right)^2}\right|_{\theta=0}=\left.\frac{\xi^2}{\mu N} \left(\frac{\langle S_x\rangle}{{\cal A}/2}\right)^2\right|_{\theta=0}
=\frac{\xi^2}{\mu N}\left(\sum_{i=1}^M |c_i|\right)^2,
\label{eq:USSS}
\end{align}
where $\xi^2=N\mbox{Var}(S_z)\big{/}|\langle S_x \rangle|^2$ is the initial squeezing parameter. Note that the uncertainty in the estimated linear combination (\ref{eq:USSS}) depends on the combination through $\sum_{i=1}^M |c_i|$.

\subsection{Comparison with a coherent spin state}
To calculate the quantum gain of this scheme, we compare it to the estimation using a coherent spin state (CSS). Since the atoms are independent, the parameters are always estimated independently and 
\begin{align}
\mbox{Var}(\hat{\cal C})_{\rm CSS}=\sum_{k=1}^Mc_k^2 \mbox{Var}(\theta_k)_{\rm CSS}\,.
\end{align}
Moreover, the uncertainty on each local parameter $\theta_k$ is always given by 
$\mbox{Var}(\theta_k)_{\rm CSS}=1/\mu_k N_k$, where $N_k$ is the number of atoms and $\mu_k$ the number of repetitions used to estimate $\theta_k$, and $\sum_k \mu_k = \mu$. 
For example, to estimate the symmetric linear combination ${\cal C}_+=\sum_{k=1}^M\theta_k/\sqrt{M}$ one can for each parameter $\theta_k$ either use $N/M$ atoms $\mu$ times or $N$ atoms $\mu/M$ times. In this sense, we can say that we have as a ``total resource" $\mu N$ independent atoms and the constraint is
\begin{align}
\sum_{k=1}^M \mu_k N_k=\mu N \,.
\end{align}
When applied to the combination ${\cal C}$, this leads to an uncertainty
\begin{align}\label{eq: Var_CSS}
\mbox{Var}(\hat{\cal C})_{\rm CSS}=\sum_{k=1}^Mc_k^2\mbox{Var}(\theta_k)_{\rm CSS}=\sum_{k=1}^M\frac{c_k^2}{\mu_k N_k}.
\end{align}
By using a Cauchy-Schwartz inequality, one can show that the uncertainty (\ref{eq: Var_CSS}) satisfies the following inequality:
\begin{align}\label{eq: Var_CSS_Bound}
\mbox{Var}(\hat{\cal C})_{\rm CSS}\geq\frac{1}{\mu N}\left(\sum_{i=1}^M |c_i|\right)^2.
\end{align}
To show it, we introduce two $M$-dimensional vectors $\bm \alpha$ and $\bm \beta$ with respective components 
\begin{align}
\alpha_k\equiv\frac{|c_k|}{\sqrt{\mu_k N_k}}\quad\textrm{and}\quad\beta_k\equiv\frac{\sqrt{\mu_k N_k}}{\sqrt{\mu N}}
\quad\textrm{with}\quad ||\bm \beta||=1.
\end{align}
Applying the Cauchy-Schwartz inequality to vectors $\bm \alpha$ and $\bm \beta$ results in
\begin{align}\label{eq: CS_Inequality}
||\bm \alpha||^2||\bm \beta||^2=\sum_{k=1}^M\frac{c_k^2}{\mu_k N_k}=\mbox{Var}(\hat{\cal C})_{\rm CSS}\geq |{\bm \alpha} \cdot {\bm \beta} |^2=\frac{1}{\mu N}\left(\sum_{i=1}^M |c_i|\right)^2.
\end{align}
This indicates that the minimal uncertainty in estimating a given linear combination ${\cal C}$ achieved by a coherent spin state is given by
\begin{align}
\label{eq: genericSQL}
\mbox{Var}(\hat{\cal C})_{\rm CSS, min}=\frac{1}{\mu N}\left(\sum_{i=1}^M |c_i|\right)^2. 
\end{align}
This minimal uncertainty is attained by the choice of $\mu_k N_k$ that saturate the Cauchy-Schwartz inequality (\ref{eq: CS_Inequality}),
\begin{equation}\label{eq: Norm_B}
\mu_k N_k= {\cal B} |c_k| \quad \mbox{with}  \quad {\cal B}=\frac{\mu N}{\sum_j |c_j| } \quad \mbox{so that} \quad \sum_{k=1}^M \mu_k N_k = \mu N \,.
\end{equation}

The quantum gain associated with the estimation of the linear combination ${\cal C}$ with a squeezed spin state compared to a coherent spin state is thus given, independently of the combination and independently of the normalization condition chosen, by
\begin{align}
\frac{\mbox{Var}(\hat{\cal C})_{\rm SSS}}{\mbox{Var}(\hat{\cal C})_{\rm CSS, min}}=\xi^2.
\end{align}

\subsection{Scanning microscope strategy}
In order to estimate the linear combination ${\cal C}$, another possibility is to use the scanning microscope (SM) strategy. In this strategy, where the local parameters are estimated consecutively and therefore independently, one has
\begin{align}
\mbox{Var}(\hat{\cal C})_{\rm SM}=\sum_{k=1}^Mc_k^2\mbox{Var}(\theta_k)_{\rm SM}.
\end{align}
For each preparation we use $N$ atoms to measure a single parameter, but we can dedicate a different number $\mu_k$ of preparations for different parameters, with
\begin{align}
\sum_{k=1}^M \mu_k=\mu \,.
\end{align}
The uncertainty on each local parameter $\theta_k$ is then
$\mbox{Var}(\theta_k)_{\rm SM}=\xi^2/\mu_k N$. 
When applied to the combination ${\cal C}$, this leads to an uncertainty
\begin{align}\label{eq: Var_SM}
\mbox{Var}(\hat{\cal C})_{\rm SM}=\sum_{k=1}^Mc_k^2\mbox{Var}(\theta_k)_{\rm SM}=\sum_{k=1}^M\frac{c_k^2 \xi^2}{\mu_k N}.
\end{align}
By using a Cauchy-Schwartz inequality, one can show that the uncertainty (\ref{eq: Var_SM}) satisfies the following inequality:
\begin{align}\label{eq: Var_SM_Bound}
\mbox{Var}(\hat{\cal C})_{\rm SM}\geq\frac{\xi^2}{\mu N}\left(\sum_{i=1}^M |c_i|\right)^2.
\end{align}
To show it, we introduce two $M$-dimensional vectors $\bm\alpha$ and $\bm\beta$ with respective components 
\begin{align}
\alpha_k\equiv\frac{|c_k|\xi }{\sqrt{\mu_k N}}\quad\textrm{and}\quad\beta_k\equiv\frac{\sqrt{\mu_k}}{\sqrt{\mu}}
\quad\textrm{with}\quad ||\bm \beta||=1 .
\end{align}
Applying the Cauchy-Schwartz inequality to vectors $\bm\alpha$ and $\bm\beta$ results in
\begin{align}\label{eq: SM_Inequality}
||\bm\alpha||^2||\bm\beta||^2=\sum_{k=1}^M\frac{c_k^2 \xi^2}{\mu_k N}=\mbox{Var}(\hat{\cal C})_{\rm SM}\geq |\bm\alpha\cdot\bm\beta|^2=\frac{\xi^2}{\mu N}\left(\sum_{i=1}^M |c_i|\right)^2.
\end{align}
This indicates that the minimal uncertainty in estimating a given linear combination ${\cal C}$ achieved by the scanning microscope is given by
\begin{align}
\mbox{Var}(\hat{\cal C})_{\rm SM, min}=\frac{\xi^2}{\mu N}\left(\sum_{i=1}^M |c_i|\right)^2. 
\end{align}
This minimal uncertainty is attained by the choice of $\mu_k$ that saturate the Cauchy-Schwartz inequality (\ref{eq: SM_Inequality}),
\begin{equation}\label{eq: Norm_D}
\mu_k = {\cal D} |c_k| \quad \mbox{with}  \quad {\cal D}=\frac{\mu}{\sum_j |c_j| } \quad \mbox{so that} \quad \sum_{k=1}^M \mu_k = \mu \,.
\end{equation}
This means that, as for the distributed entanglement strategy, for the scanning microscope strategy the quantum gain of the squeezed spin state with respect to a coherent spin state is $\xi^2$, independently of the combination and independently of the normalization condition chosen. However, we should point out a limitation of the scanning microscope strategy that necessitates the estimation of all parameters and thereby requires a number of system preparations $\mu$ that is at least equal to the number of sensors $M$. As a result, if $\mu< M$, this strategy  cannot be applied for estimating a given linear combination of the parameters.

\section{Experimental sequence}
\begin{figure}
\centering
{\includegraphics[width=1\linewidth]{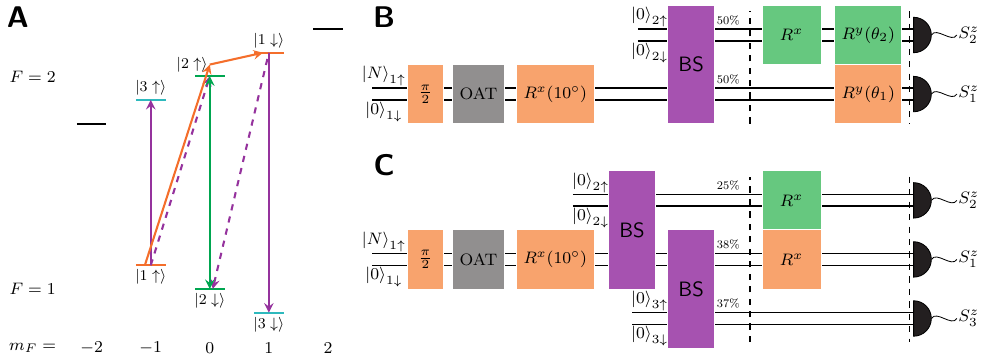}}
\caption{\label{fig6} (A) $^{87}$Rb ground state hyperfine structure with Zeeman levels $\ket{F, m_F}$. Levels used in our experiment are shown in color, consistent with the color code in (B). Purple arrows indicate microwave transitions that are used to coherently split the cloud. Green and orange arrows indicate the single-photon transition used to rotate $\mathbf{S}_2$ and the two-photon transition used to rotate $\mathbf{S}_1$, respectively. (B) and (C) show the experimental sequence for two-parameter and three-parameter estimation, respectively. Three main parts of the sequence are separated by dashed lines: preparation of the atomic sensor array (left), spin rotations of the individual sensors (middle), and individual detection (right). The BEC is initially prepared in a superposition of $\ket{\Au}$ and $\ket{\Ad}$ by the first $\pi/2$ pulse. A one-axis twisting (OAT) Hamiltonian is applied to generate a global spin-squeezed state, followed by a small rotation of $10^\circ$ around the $x$-axis to align the squeezed spin component with the $z$-axis. Subsequently, an atomic beam splitting sequence (BS) coherently splits the initial BEC into an array of atomic sensors. To adjust the correlations between sensors, local $\pi$ rotations around the $x$-axis $R^x$ can be performed on certain sensors if desired. Local parameters $\theta_k$ can then be encoded on the $k$-th sensor through small rotations around the $y$-axis $R^y(\theta_k)$. Finally, absorption images are taken to measure $S^z_{k}$ for all sensors. 
}
\end{figure}

The experimental sequence is illustrated in Fig.~\ref{fig6}. We start from a BEC with $N\approx 1450$ atoms on an atom chip \cite{Boehi2009}, initially prepared in state $\ket{F=1, m_F=-1}$. A first $\pi/2$ pulse transfers the atoms into an equal superposition of $\ket{\Au}\equiv \ket{F=1, m_F=-1}$ and $\ket{\Ad}\equiv \ket{F=2, m_F=1}$, defining the collective spin $\mathbf{S}_1$. In this coherent spin state (CSS) the atoms are uncorrelated. To entangle the atoms we harness atomic collisions in a spin-dependent potential to apply a one-axis twisting Hamiltonian~\cite{Riedel2010,Ockeloen2013} for a well-defined time. The resulting squeezed spin state points in what we define as the $x$-direction. It is then rotated around the $x$-axis such that it shows minimal variance in $S^z$, which is reduced by $\xi^2=-6.5(2)$~dB compared to the SQL. 

In order to spatially split the BEC into $M=2$ atomic sensors, we release  the atoms from the trap and simultaneously couple the states $\ket{\Au}$ and $\ket{\Ad}$ to the states $\ket{\Bu}\equiv \ket{F=2, m_F=0}$ and $\ket{\Bd}\equiv \ket{F=1, m_F=0}$, respectively, which define the collective spin $\mathbf{S}_2$. The coupling is done by driving the two transitions with a two-tone microwave (MW) magnetic field. The duration of the coupling pulse determines the atom numbers $N_1$ and $N_2$ in the two sensors, which we choose to be equal for the experiments in Figs.~\ref{fig:correlations} and \ref{fig:jointestimate} of the main paper, while we adjust them to different values for the experiments shown in Fig.~\ref{fig:UnequalSplit} of the main paper. 

Exploiting the difference in magnetic moments of the involved states, the atomic ensembles realizing the spins $\mathbf{S}_1$ and $\mathbf{S}_2$ are then spatially separated by applying a magnetic field gradient pulse, resulting in two atomic clouds separated by more than $14~\mu$m at this point in the sequence \cite{Colciaghi2023}. The spin lengths are $\mean{S_k^x}=CN_k/2$, where $C$ corresponds to the interferometric contrast in a Ramsey measurement;  independent measurements show $C\approx 94\%$ for both spins in our experiment. To prepare the different sensor configurations, we either apply a spin rotation $R^x$ of $\mathbf{S}_2$ around the $x$-axis by an angle $\pi$, or we omit this pulse. Subsequently, we encode the parameters $\theta_{1}$ and $\theta_{2}$ on each sensor as small angle rotations around the $y$-axis. This translates the local parameters into an observable change in $S_{1}^z$ and $S_{2}^z$, respectively. 

After the parameter encoding, we take absorption images to measure the atom numbers $N_{\Au}$, $N_{\Ad}$, $N_{\Bu}$, $N_{\Bd}$ in all four states, corresponding to simultaneous measurements of both $S_1^z=(N_{\Au}-N_{\Ad})/2$ and $S_2^z=(N_{\Bu}-N_{\Bd})/2$, from which we obtain $\theta_1 \approx S_1^z / \langle S_1^x\rangle$ and $\theta_2 \approx S_2^z / \langle S_2^x\rangle$ using $\theta_k\ll 1$. 

To demonstrate our ability to encode varying local parameters on both spins $\mathbf{S}_1$ and $\mathbf{S}_2$ we have taken many datasets where we vary both parameters. For the data presented in Figs. \ref{fig:correlations} and \ref{fig:jointestimate} we chose a dataset where $\mathbf{S}_1$ was subjected to a rotation of $-5^\circ$ and varying rotations of $-10^\circ$, $-5^\circ$, $0^\circ$, $5^\circ$ and $10^\circ$ were applied to $\mathbf{S}_2$. Unbalanced atom loss and small miscalibrations of the phases of the MW pulses lead to small offsets of $S_{1}^z$ and $S_{2}^z$ even when the spins are not rotated before detection. We subtract these constant offsets from the data, so that $\theta_{1}=\theta_{2}=0$ corresponds to the center of the dataset, to simplify the presentation. For the stated quantum gain and parameter variances these offsets do not play any role. 

Technical noise in our experiment increases all measured variances. The major source of technical noise is photon shot noise on the absorption images. To reduce this source of noise we optimize our imaging parameters and select an elliptical region of interest when counting the atom number on the absorption images, see ref.~\cite{Colciaghi2023} (Appendix C). Although this technical noise can be accurately measured independently, we do not subtract it or any other technical noise when processing the data. 

An important aspect in stating quantum enhanced measurement precision is an accurate calibration of the resources used. In particular the atom number has to be calibrated accurately. In our experiment, we calibrate the detected atom numbers in the absorption images with the method described in \cite{muessel2013}. Additionally, we perform the experiment with independent (unentangled) sensors by using the same sequence but excluding the one-axis twisting evolution that generates the entanglement. The obtained data are analyzed in the exact same way, and show a variance of $0.5(3)$~dB above the SQL, see the gray points in Fig.~\ref{fig:jointestimate}~C,~D of the main paper.

In order to prepare an array of $M=3$ atomic sensors, we use the same techniques as described above, but additionally make use of the Zeeman states $\ket{3\uparrow}\equiv \ket{F=2,m_F=-1}$ and $\ket{3\downarrow}\equiv \ket{F=1,m_F=1}$, which define $\mathbf{S}_3$. We first distribute the atoms between $\mathbf{S}_1$ and $\mathbf{S}_2$ with a two-tone microwave pulse, as described above. Subsequently, another two-tone MW pulse drives the transitions indicated by the purple arrows in Fig.~\ref{fig6}A, transferring atoms from  $\ket{1\uparrow}$ and $\ket{1\downarrow}$ to $  \ket{3\uparrow}$ and  $\ket{3\downarrow}$, respectively. This realizes two sequential atomic beamsplitters. The duration of the two pulses determines $N_1$, $N_2$, and $N_3$. The three atomic ensembles realizing the spins $\mathbf{S}_1$, $\mathbf{S}_2$, and $\mathbf{S}_3$ are then spatially separated by applying a magnetic field gradient pulse, again making use of the difference of the magnetic moments of the involved states. Before the spins are detected by absorption imaging, $\pi$ rotations of $\mathbf{S}_1$ and $\mathbf{S}_2$ are applied if desired to adjust the correlations among the three sensors and realize the different measurement configurations. In the three sensor case, the atoms are distributed to more clouds on the absorption images, which leads to an increase of the overall detection noise.  We therefore increased the total number of atoms in these 
experiments to $N\approx  1700$ to maintain a good signal-to-noise ratio.

\section{Data analysis}

The precision gain in joint multiparameter estimation results from the correlations between measurement results of the individual sensors. In harnessing these correlations, certain linear combinations of parameters play an important role. In the case of symmetric splitting of an ideal spin-squeezed state into two sensors, these linear combinations are $\theta_+$ and $\theta_-$, which make optimal use of the correlations and also allow for an intuitive explanation of how the precision is gained, as detailed in the main text. More formally, this is directly linked to a certain form of the covariance matrices describing the individual measurement configurations. In the particular case considered here, i.e.\ configuration $1$ without and configuration $2$ with $\pi$ pulse applied to $\mathbf{S}_2$, respectively, the covariance matrices must have the form
\begin{align}
\Sigma^{(\mu_1)}_1=\left(
\begin{matrix} 
v&c\\c&v    
\end{matrix}\right)\qquad \textrm{and}\qquad \Sigma^{(\mu_2)}_2=\left(
\begin{matrix} 
v&-c\\-c&v    
\end{matrix}\right)
\end{align}  
with $v\geq|c|$. This is precisely the form obtained by symmetric splitting of an ideal spin squeezed state, see Eqs.~(\ref{eq: SigmaDiag}) and (\ref{eq: SigmaHorDiag}). 
In an experiment, however, finite statistics due to a finite number of  measurements or different technical noise in different configurations lead to deviations from this idealized scenario.  In the experiment one is therefore tasked to find an optimal estimator $\hat{\theta}_k$ for the local parameter $\theta_k$  which is a linear combination of the local measurement results of all configurations. The only constraint for this linear combination comes from the requirement that the estimator is unbiased, i.e. $\mean{\hat{\theta}_k}=\theta_k$. This constraint can be formulated in general for the estimation of the parameter combination $\mathbf{n}\cdot\boldsymbol{\theta}$ as

\begin{align}
\label{eq:estimatorrestriction}
\mathbf{n}= \sum_{\lambda} \mathbf{x}_\lambda,
\end{align}
where $\lambda$ is the index of the measurement configuration and the $M$-dimensional vectors $\mathbf{x}_\lambda$ are the coefficients for the local parameters in the configuration $\lambda$. Note that the vectors $\mathbf{x}_\lambda$ do not need to be normalized, and their norm can be regarded as a weight for the respective configuration. The variance of this estimator is 
\begin{align}
\label{eq:varestimator}
\Var(\mathbf{n}\cdot\boldsymbol{\theta})= \sum_{\lambda} \mathbf{x}_\lambda^{T}\Sigma_\lambda\mathbf{x}_\lambda,
\end{align}
where $\Sigma^{(\mu_\lambda)}_\lambda$ is $\mu_\lambda^{-1}$ times the $M\times M$-dimensional sample covariance matrix of the $\mu_\lambda$ measurement results obtained in configuration $\lambda$. If one is interested in the local parameter $\theta_k$, one has to choose $\mathbf{n}$ as the $k$th standard basis vector. 
In this general case we find the coefficient vectors $\mathbf{x}_\lambda$ for each local parameter numerically by minimizing Eq.~(\ref{eq:varestimator}) under the constraint in Eq.~(\ref{eq:estimatorrestriction}).  

\subsection*{Data analysis for $M=2$}
In the case of $M=2$ sensors our experimental results are actually very close to the ideal symmetric case discussed above. An analysis of the data in terms of $\theta_+$ and $\theta_-$ as described in the main text corresponds to fixing $\mathbf{x}_1=\frac{1}{2}\left(\begin{matrix}1\\ \alpha_1\end{matrix}\right)$ and  $\mathbf{x}_2=\frac{1}{2}\left(\begin{matrix}1\\ -\alpha_1\end{matrix}\right)$ for the estimation of $\theta_1$, and $\mathbf{x}_1=\frac{1}{2}\left(\begin{matrix}\alpha_2\\ 1\end{matrix}\right)$ and  $\mathbf{x}_2=\frac{1}{2}\left(\begin{matrix}-\alpha_2\\ 1\end{matrix}\right)$ for the estimation of $\theta_2$, where $\alpha_1$ and $\alpha_2$ are numerically optimized and correspond to the weight in Eq.~\ref{eq: Estimator}. For the dataset with $\theta_1=\theta_2=0$, this results in a quantum gain of -3.55(15)~dB and -3.37(12)~dB for $\theta_1$ and $\theta_2$. If instead we use the more general optimization strategy described above, where the only constraint on $\mathbf{x}_\lambda$ is given by Eq.~(\ref{eq:estimatorrestriction}), we obtain -3.63(17)~dB and -3.47(11)~dB for the two local parameters. As expected, the more general optimization yields slightly better results. For all data presented in the main paper we use the more general optimization strategy.

In the experiment presented in Fig.~\ref{fig:jointestimate} of the main paper, we encode five different parameter values for $\theta_2$. The whole dataset contains 11200 measurements and was taken within 90 hours of measurement time. For each encoded $\theta_2$, we perform $\mu\approx 2200$ measurements, where we alternate between the two configurations (with or without the $\pi$ rotation of $\mathbf{S_2}$). We analyze the data in blocks of 200 sequential measurements and obtain the variance and quantum gain of each block. In addition, for each block 20 measurements of the Ramsey contrast $C$ are taken. The data presented in the main text are the mean and standard error of these blocks. This way of grouping the data into blocks renders the analysis robust against slow drifts of the experimental conditions. However, we want to emphasize that the subdivision into blocks only affects the quantitative results slightly and does not change any of the qualitative statements made.

To obtain the histograms shown in Fig.~\ref{fig:jointestimate}D, we combine pairs of consecutive measurements in the two configurations to obtain the local parameters using the nonlocal entanglement. 

In the experiment presented in Fig.~\ref{fig:UnequalSplit}, we drive the splitting pulse for different durations to adjust the relative size of the two sensor spins, and we take measurements with $\frac{N_1}{N}=0.81,0.71,0.64,0.52,0.40$. Together with the $\pi$ spin rotation, we obtain a total of ten different configurations. For each of these datasets, we numerically determine the optimal linear combination $\mathbf{n}\cdot\boldsymbol{\theta}=\cos (\alpha)\,\theta_1 + \sin (\alpha)\,\theta_2$ that maximizes the quantum gain. The SQL 
for independent atoms in this scenario is ${(\cos(\alpha)+\sin(\alpha))^2}/{N}$, see Eq.~(\ref{eq: genericSQL}). The quantum gain is then defined as ${N \Var(\mathbf{n}\cdot\boldsymbol{\theta})}/{(\cos(\alpha)+\sin(\alpha))^2}$. 

\subsection*{Data analysis for $M=3$}
In the experiment with $M=3$ sensors, the distribution of atoms $N_k$ is not symmetrical, due to an inadvertent miscalibration of a microwave pulse. For the joint estimation, we therefore only use the general optimization of the local estimators, which can be applied for any distribution of $N_k$. In total this dataset includes 1578 measurements, where each of the four configurations was repeated approximately 400 times and it was taken within 13 hours. The stated uncertainties on the variances are standard errors, which we calculate as relative errors  $\sigma(\Var)/\Var=\sqrt{2/(\mu-\mathrm{d.o.f.})}$ where $\mu$ is the number of measurements and $\mathrm{d.o.f.}$ are the degrees of freedom, i.e. the number of free parameters, used for optimizing the variance. In the simplest case, $\mathrm{d.o.f.}=1$  accounts for the fact that the sample mean is estimated from the sample. In our case of four configurations with three sensors,  three out of the twelve coefficients in the $\mathbf{x}_\lambda$ are fixed by the constraint in Eq.~(\ref{eq:estimatorrestriction}) bringing the total 
to $\mathrm{d.o.f.}=10$. The contrasts for $\mathbf{S_1}$ and $\mathbf{S_2}$ are $C_1=0.95(1)$ and $C_2=0.90(1)$ and the contrast of $\mathbf{S_3}$ is estimated as $C_3= 0.95$, consistent with the other measurements.

\end{document}